\documentclass[twocolumn]{aastex631}

\usepackage{amsmath}
\usepackage{subfigure}
\usepackage{xcolor}
\usepackage[T1]{fontenc}
\usepackage[normalem]{ulem}
\usepackage{chngcntr}

\shorttitle{M dwarf planets}
\shortauthors{Chachan \& Lee} 

\begin{document}

\title{Small Planets Around Cool Dwarfs: \\
Enhanced Formation Efficiency of Super-Earths around M dwarfs}

\author[0000-0003-1728-8269]{Yayaati Chachan}
\affiliation{Department of Physics and Trottier Space Institute, McGill University, 3600 rue University, H3A 2T8 Montreal QC, Canada}
\affiliation{Trottier Institute for Research on Exoplanets (iREx), Universit\'e de Montr\'eal, Canada}

\author[0000-0002-1228-9820]{Eve J. Lee}
\affiliation{Department of Physics and Trottier Space Institute, McGill University, 3600 rue University, H3A 2T8 Montreal QC, Canada}
\affiliation{Trottier Institute for Research on Exoplanets (iREx), Universit\'e de Montr\'eal, Canada}

\correspondingauthor{Yayaati Chachan}
\email{yayaati.chachan@mcgill.ca}

\begin{abstract}

Current measurements of planet population as a function of stellar mass show three seemingly contradictory signatures: close-in super-Earths are more prevalent around M dwarfs than FGK dwarfs; inner super-Earths are correlated with outer giants; and outer giants are less common around M dwarfs than FGK dwarfs. Here, we build a simple framework that combines the theory of pebble accretion with the measurements of dust masses in protoplanetary disks to reconcile all three observations. First, we show that cooler stars are more efficient at converting pebbles into planetary cores at short orbital periods. Second, when disks are massive enough to nucleate a heavy core at 5 AU, more than enough dust can drift in to assemble inner planets, establishing the correlation between inner planets and outer giants. Finally, while stars of varying masses are similarly capable of converting pebbles into cores at long orbital periods, hotter stars are much more likely to harbor more massive dust disks so that the giant planet occurrence rate rises around hotter stars. Our results are valid over a wide range of parameter space for a disk accretion rate that follows $\dot{M}_\star \sim 10^{-8}\,M_\odot\,{\rm yr}^{-1}(M_\star/M_\odot)^2$. We predict a decline in mini-Neptune population (but not necessarily terrestrial planets) around stars lighter than $\sim$0.3--0.5$M_\odot$. Cold giants ($\gtrsim$5 AU), if they exist, should remain correlated with inner planets even around lower mass stars.

\end{abstract}

\section{Introduction} \label{sec:intro}
Super-Earths and mini-Neptunes are more likely to appear around M dwarfs, with their occurrence rate inside $\sim$50--100 days enhanced by factors of $\sim$3--10 as compared to those around FGK dwarfs \citep[e.g.,][]{Dressing15, Mulders15, Gaidos2016, Hsu20}. One way to explain this difference is to consider the lower binarity of M dwarfs compared to their more massive counterparts \citep[e.g.,][]{Raghavan10} as wide binaries can spuriously amplify the stellar brightness and dilute transit signals. Correcting for the effects of binaries however can only account for roughly half of the measured difference \citep{Moe21}.

Another difference between planetary systems around M vs.~FGK dwarfs is the existence of gas giants. Unlike small planet statistics, gas giant occurrence rate of FGK dwarfs is boosted by factors of $\sim$2--3 as compared to M dwarfs \citep[e.g.,][]{Clanton14,Fulton21}. This general trend of gas giant appearing more commonly around more massive stars is corroborated in direct imaging surveys as well with a caveat that those surveys are sensitive to super-Jupiters around young stars \citep{Nielsen19}. Gas giants, when they exist, are much more likely to appear at intermediate distances $\sim$1--10 AU \citep{Fulton21}. Recently, \citet{Mulders21} argued that this outer giant acts as a barrier against the radial drift of pebbles which can effectively starve the inner disk from solids and thereby limiting the amount of pebbles available for the formation of planetary cores \citep[see also][]{Lambrechts19}.

Such hypothesis runs counter to the observations that show a correlation between outer gas giants and inner super-Earths \citep[][]{Zhu18, Bryan19, Herman19, Rosenthal22}.\footnote{Recent radial velocity follow-up to Kepler/K2 \citep{Bonomo2023} and TESS planets \citep{VanZandt2023} report that giant planet occurrence rate around stars with inner planets is statistically consistent with that of field stars. Both studies call to attention the need for a larger sample so at present, we conclude that there is either a neutral or positive correlation between outer gas giants and inner super-Earths.} In fact, the core assembly process by pebble accretion is notoriously lossy \citep[e.g.,][]{Ormel17, Ormel18}, so much so that for every pebble isolation mass $\sim$10$M_\oplus$ that is created at $\sim$10 AU, upwards of $\sim$100$M_\oplus$ worth of material can be drifted into the inner orbits while (but not after) the massive core assembles \citep{Lin18, Chachan22}. Furthermore, once the giant forms, the interplay between its secular resonance with planetesimals and aerodynamic drag could spell rapid transport of the latter, enriching the inner region with even more solid material \citep{Best23}.

In this work, we build a simple framework that combines the theory of pebble accretion with the measured distribution of protoplanetary disk masses to reconcile all three observations: the propensity of cooler stars to harbor more super-Earths; the correlation between inner super-Earths and outer giants; and the tendency for cooler stars to harbor less giants. We lay out the underlying theory in Section \ref{sec:theory}. Results are presented in Section \ref{sec:results} and we provide a summary and predictions in Section \ref{sec:discussion}.

\section{Theory}
\label{sec:theory}
We begin by defining
the efficiency $\epsilon$ with which a protoplanet accretes pebbles:
\begin{equation}
    \epsilon = \frac{\dot{M}_{\rm peb}}{\dot{M}_{\rm drift}},
\end{equation}
where $\dot{M}_{\rm peb}$ is the accretion rate of pebbles and $\dot{M}_{\rm drift}$ is the radial drift rate of pebbles at the protoplanet's location. We assume $\dot{M}_{\rm drift} = 2\pi r v_{\rm r} \Sigma_{\rm d, St}$, where $r$ is the cylindrical distance from the star, $\Sigma_{\rm d, St}$ is the surface density of dust grains of Stokes number St and $v_{\rm r}$ is their radial velocity (negative sign denotes inward drift),
\begin{align}
    v_{\rm r} &= -\frac{3}{2}\frac{\nu}{r}\frac{1}{1+{\rm St}^2} - 2\eta v_{\rm K} \frac{\rm St}{1+{\rm St}^2} \nonumber \\
    &= -\frac{3}{2} \frac{c_s^2}{v_k}\frac{1}{1+{\rm St}^2} \left[\alpha_{\rm t} + \frac{2}{3}|\gamma|{\rm St}\right]
    \label{eq:vr_dust}
\end{align}
where the first term accounts for the particles' coupling to the inward accretion of gas: $\nu \equiv \alpha_{\rm t} c_{\rm s} H_{\rm g}$ is the kinematic viscosity of the gas, $\alpha_{\rm t}$ is the Shakura-Sunyaev parameter, $c_{\rm s} = \sqrt{k_{\rm B} T_{\rm disk} / \mu m_{\rm H}}$ is the sound speed, $k_{\rm B}$ is the Boltzmann constant, $T_{\rm disk}$ is the local disk temperature, $\mu = 2.3$ is the mean molecular weight of the gas, $m_{\rm H}$ is the mass of a hydrogen atom, $H_{\rm g} \equiv c_{\rm s} / \Omega_{\rm K}$ is the gas disk scale height, $\Omega_{\rm K} = \sqrt{GM_\star/r^3}$ is the Keplerian orbital frequency, $G$ is the gravitational constant, and $M_\star$ is the mass of the central star. The second term of equation \ref{eq:vr_dust} accounts for the Nakagawa-Sekiya-Hayashi drift velocity from the gas headwind \citep{Nakagawa86}: $\eta \equiv -0.5 \gamma (c_{\rm s} / v_{\rm K})^2$ is a measure of the deviation in gas orbital velocity from the Keplerian velocity $v_{\rm K} = r \Omega_{\rm K}$ due to the radial pressure gradient in the gas disk with $\gamma \equiv {\rm d \, ln}P_{\rm g} / {\rm d \, ln}r$ and $P_{\rm g}$ is the gas pressure. We limit our analysis to ${\rm St} \lesssim 1$ so that $(1+{\rm St}^2)^{-1} \rightarrow 1$ within order unity. 

Following \citet[][see also \citealt{Ormel10, Ormel17}]{Lin18}, the rate of pebble accretion can be written as
\begin{equation}
    \dot{M}_{\rm peb} = 2\Sigma_{\rm d, St} R_{\rm acc} v_{\rm acc} \times {\rm min}(1, R_{\rm acc}/H_d)
\end{equation}
where particles that come within $R_{\rm acc}$ of the protoplanet at speed $v_{\rm acc}$ will be accreted onto the core and $H_d = H_{\rm g}[\alpha_t/(\alpha_t+{\rm St})]^{1/2}$ is the pebble disk scale height. When $R_{\rm acc} < H_d$, we are in three-dimensional (3D) accretion regime and under the condition that the pebble-protoplanet encounter time is longer than the settling time, the terminal velocity of the pebbles is approximately a quarter of the accretion velocity \citep{Ormel10}, which allows us to write $R_{\rm acc}^2 v_{\rm acc} \sim 4 G M_{\rm p} {\rm St} / \Omega_{\rm K}$. The accretion rate in the 3D regime is therefore given by
\begin{equation}
    \dot{M}_{\rm peb} \sim \frac{8\Sigma_{\rm d, St}GM_{\rm p}{\rm St}}{c_s}\left(\frac{\alpha_{\rm t} + {\rm St}}{\alpha_{\rm t}}\right)^{1/2}
\end{equation}
where $M_{\rm p}$ is the mass of the protoplanet. When $R_{\rm acc} > H_d$ (two-dimensional accretion) and $v_{\rm acc}$ is dominated by the headwind velocity $v_{\rm hw} = \eta v_{\rm K}$ (2D, hw),
\begin{equation}
    \dot{M}_{\rm peb} \sim 4\eta^{1/2}{\rm St}^{1/2}q^{1/2}\Omega_{\rm K}\Sigma_{\rm d, St}r^2,
\end{equation}
where $q \equiv M_{\rm p}/M_\star$. This condition is satisfied when (see the appendix of \citealt{Lin18}; see also \citealt{Perets11})
\begin{equation}
    {\rm St}_{\rm 2D, hw} > \dfrac{1}{2 \sqrt{2}} \alpha^{1/2} |\gamma|^{1/2} q^{-1/2}  \bigg(\dfrac{c_{\rm s}}{v_{\rm K}}\bigg)^2 .
    \label{eq:St_2D_hw}
\end{equation}
In the same two-dimensional accretion but when $v_{\rm acc} \sim 3 \Omega_{\rm K} R_{\rm acc} / 2$ is dominated by Keplerian shear  (2D, sh),
\begin{equation}
    \dot{M}_{\rm peb} \sim (192)^{1/3}{\rm St}^{2/3}q^{2/3}\Omega_{\rm K}\Sigma_{\rm d, St}r^2.
\end{equation}
whose condition is satisfied when (equation A5 of \citealt{Lin18})
\begin{equation}
    {\rm St}_{\rm 2D, sh} > \alpha_{\rm t}^{3/5} \bigg(\dfrac{3}{8}\bigg)^{2/5} \bigg(\dfrac{c_{\rm s}}{v_{\rm K}}\bigg)^{6/5} q^{-2/5}.
    \label{eq:St_2D_sh}
\end{equation}
The two-dimensional accretion transitions from headwind- to shear-dominated regime when $M_{\rm p}$ exceeds a transition mass
\begin{equation}
    M_p \gtrsim \frac{M_t}{8{\rm St}} = \frac{(\eta v_{\rm K})^3}{8G\Omega_{\rm K}}{\rm St}^{-1};
\end{equation}
equivalently,
\begin{equation}
    {\rm St}_{\rm 2D} > \dfrac{1}{64} |\gamma|^3  \bigg(\dfrac{c_{\rm s}}{v_{\rm K}}\bigg)^6 q^{-1}.
\end{equation}
Combining everything together,
\begin{align}
&\epsilon= \nonumber \\
&\begin{cases} 
    \dfrac{2}{3\pi}\bigg(\dfrac{q{\rm St}}{\eta}\bigg)^{1/2} |\gamma|\bigg(\alpha_t+\dfrac{2}{3}|\gamma|{\rm St}\bigg)^{-1} & \hspace{.1cm} (\text{2D, hw})\\
    \dfrac{(192)^{1/3}}{6\pi}\bigg(q{\rm St}\bigg)^{2/3}\bigg(\dfrac{|\gamma|}{\eta}\bigg)\bigg(\alpha_t+\dfrac{2}{3}|\gamma|{\rm St}\bigg)^{-1} & \hspace{.1cm} (\text{2D, sh})\\
    \dfrac{4}{3\pi}q\dfrac{r}{H_{\rm d}}\dfrac{|\gamma|{\rm St}}{\eta}\bigg(\alpha_t+\dfrac{2}{3}|\gamma|{\rm St}\bigg)^{-1}. & \hspace{.1cm} (\text{3D})
\end{cases}
\label{eq:acc-eff}
\end{align} 

We define a mass averaged efficiency $\bar{\epsilon}$ by integrating over a grain size distribution assuming Epstein drag so that ${\rm St} \propto a$ and grain mass $m \propto {\rm St}^3$:
\begin{equation}
    \bar{\epsilon} (M_{\rm p}) = \dfrac{\int \epsilon (M_{\rm p}, {\rm St}) \, n({\rm St}) \, m({\rm St}) \, {\rm d}{\rm St}}{\int  n({\rm St}) \, m({\rm St}) \, {\rm d}{\rm St}}
    \label{eq:avg_eps}
\end{equation}
where $n({\rm St}) = dn/d{\rm St}$ is the grain size distribution such that $n({\rm St})m({\rm St}) \propto {\rm St}^{-0.5}$ for St $<2\alpha_{\rm t}/\pi$ (turbulent regime) and $n({\rm St})m({\rm St}) \propto {\rm St}^{-0.75}$ for St $>2\alpha_{\rm t}/\pi$ (settling regime) as inferred from \citet{Birnstiel11}.

The upper limit on St is set by fragmentation from turbulent velocity dispersion
\begin{equation}
    {\rm St}_{\rm f} = \frac{1}{3\alpha_{\rm t}} \frac{v_{\rm f}^2}{c_{\rm s}^2}, 
    \label{eq:st_frag}
\end{equation}
where $v_{\rm f}$ is the collision velocity at which grains fragment. Grains may drift before they reach St$_{\rm f}$ but this is not the limiting process for the vast majority of our parameter space. This effect only appears for low mass stars at $\alpha_{\rm t} = 10^{-4}$, $v_{\rm f} = 10$ m/s, and $P \gtrsim 10^3$ days, where it makes a negligible difference to our results and is therefore left out of subsequent discussion. We integrate $\epsilon$ over St $\in [10^{-6}, {\rm St}_{\rm f}]$, where the lower limit of St is chosen to have enough dynamic range while also capturing the case where the radial inflow of particles is dominated by coupling to gas inflow (${\rm St} < (3/2) \alpha_{\rm t} / |\gamma|$; see equation \ref{eq:acc-eff}).
Once we obtain $\bar{\epsilon} (M_{\rm p})$, we calculate the dust mass required for a protoplanet to reach the pebble isolation mass $M_{\rm iso}$
\begin{equation}
    M_{\rm dust \rightarrow iso} = \int_{M_0}^{M_{\rm iso}} \frac{1}{\bar{\epsilon} (M_{\rm p})} {\rm d}M_{\rm p} 
    \label{eq:mdust-to-iso}
\end{equation}
where the initial protoplanet mass $M_0$ is taken to be a lunar mass ($0.01~M_{\oplus}$; \citealt{Johansen17}). Choosing a stellar mass dependent $M_0$ \citep[e.g.,][]{Liu2020} has a negligible impact on our results (see \S~\ref{sec:analy_expect}). We limit our integration to $M_{\rm iso}$ as at this point, the planet can perturb the surrounding disk gas to create a pebble trap exterior to its orbit \citep[e.g.,][]{Lambrechts14} and so pebble accretion halts. We adopt $M_{\rm iso}$ derived by \citet{Bitsch18}, relevant for $\alpha_{\rm t} \gtrsim 10^{-4}$:

\begin{multline}
    M_{\rm iso} = 25 M_{\oplus} \bigg(\frac{M_{\star}}{M_{\odot}}\bigg) \bigg(\frac{H_{\rm g}/r}{0.05}\bigg)^3  \bigg[1 - \frac{\gamma + 2.5}{6} \bigg] \\ \times \bigg[0.34 \bigg(\frac{-3}{{\rm log_{10}} \alpha_{\rm t}}\bigg)^4 + 0.66 \bigg].
    \label{eq:pebble_iso}
\end{multline}

\begin{figure*}
    \centering
    \includegraphics[width=\textwidth]{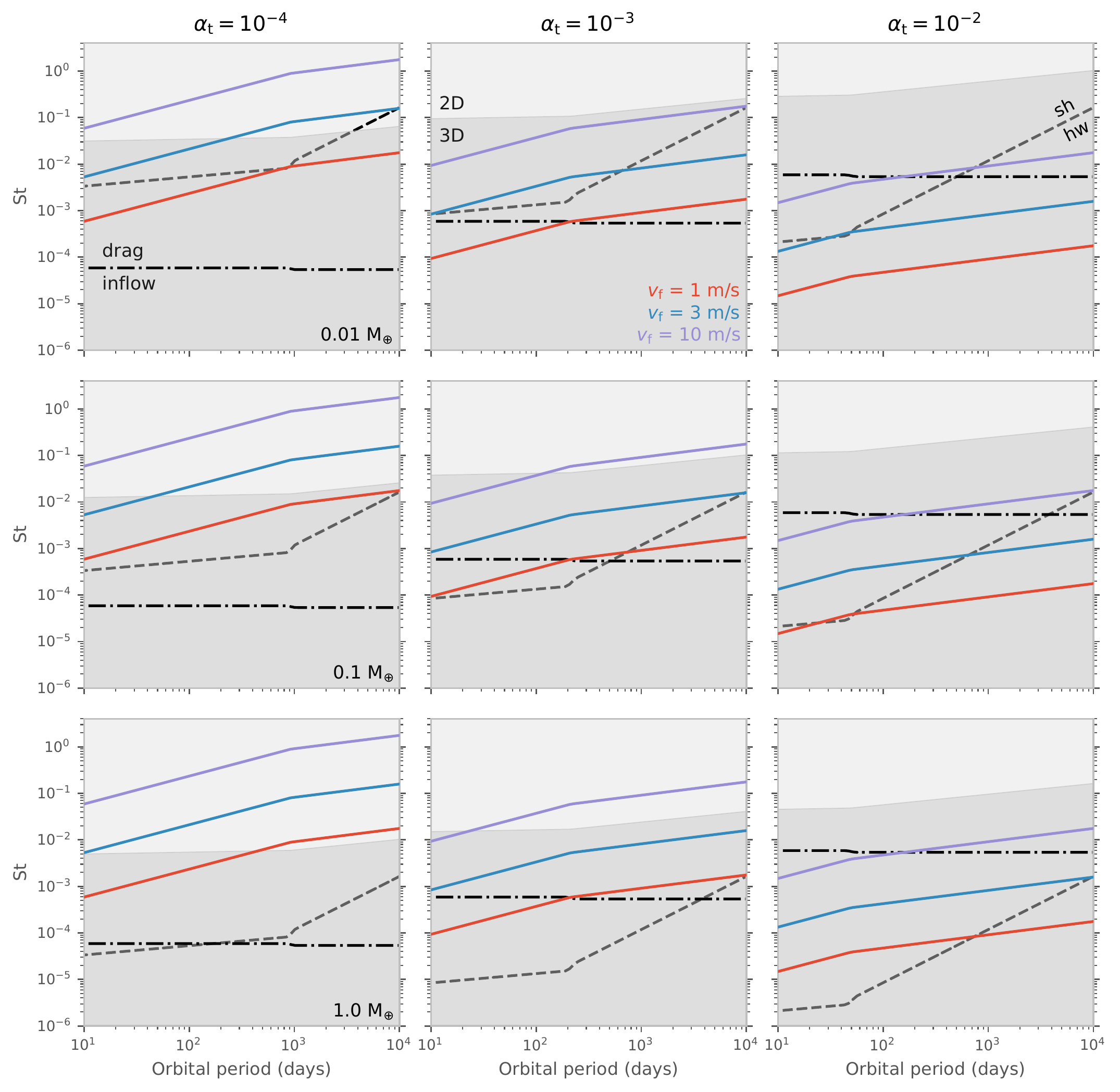}
    \caption{Stokes parameters St that delineate different accretion regimes for a 0.5$M_\odot$ host star and $\dot{M}_{1\odot}=10^{-8}\,M_\odot\,{\rm yr}^{-1}$. Different columns show $\alpha_{\rm t} = 10^{-4}$, $10^{-3}$, and $10^{-2}$ from left to right while different rows show protoplanet masses $M_{\rm p}=0.01$,0.10, and 1.00$M_\oplus$ from top to bottom. The colored curves draw ${\rm St}_{\rm f}$ and they are each labelled with corresponding fragmentation velocity $v_{\rm f}$. The minimum ${\rm St}$ required for a transition from headwind to shear-dominated accretion is drawn in black dashed line while the 3D and 2D accretion regimes are represented by different shades of grey. The transition to 2D regime is calculated assuming the accretion is already shear-dominated. This assumption fails only for our lowest protocore mass $M_{\rm p}=0.01M_\oplus$ and lowest $\alpha_{\rm t}=10^{-4}$ at wide separations ($P \gtrsim 10^4$ days). In black dot-dashed lines, we delineate the minimum ${\rm St}$ for the radial motion of the particles to be dominated by the aerodynamic drag and drift rather than coupling to the viscous gas motion, set by ${\rm St} > 3\alpha_{\rm t}/2|\gamma|$. The broken power-law feature in all ${\rm St}$-period relations reflect the switch from accretion-dominated (inner orbit) to irradiation-dominated (outer orbit) disk heating.}
    \label{fig:St_accretion}
\end{figure*}
 
We use the disk temperature profile of \cite{Ida16}, which draws from the more detailed models in \cite{Garaud07} and \cite{Oka11} and agrees with numerical radiative-transfer calculations from \cite{Dalessio98}, within order unity numerical coefficients. In the viscously heated inner regions of the disk, the temperature $T_{\rm vis}$ is given by
\begin{multline}
    T_{\rm vis} \simeq 200 \, {\rm K} \bigg(\frac{M_\star}{M_{\odot}}\bigg)^{3/10} \bigg(\frac{\dot{M}_\star}{10^{-8} M_{\odot} {\rm yr}^{-1}}\bigg)^{2/5} \\ \bigg(\frac{\alpha_{\rm t}}{10^{-3}}\bigg)^{-1/5} \bigg(\frac{r}{1 \, {\rm au}}\bigg)^{-9/10}
\end{multline}
where $\dot{M}_\star$ is the rate of accretion onto the star. Observational constraints of $\dot{M}_\star$ indicate that it increases with $M_\star$:
\begin{equation}
   \dot{M}_\star \simeq \dot{M}_{1\odot}  \bigg(\frac{M_\star}{M_{\odot}}\bigg)^B,
\end{equation}
where the power-law index $B$ ranges from 1.5 and 3.1 and $\dot{M}_{1\odot}$ is the mass accretion rate onto a solar mass star. We choose $B=2.0$ which is the measured mean $\dot{M}_\star$--$M_\star$ relationship and we vary $\dot{M}_{1\odot}$ between $10^{-9} M_{\odot} \, {\rm yr}^{-1}$ and $10^{-8} M_{\odot} \, {\rm yr}^{-1}$ which captures 30--70 percentile measured accretion rates for a solar mass star \citep[e.g.,][]{Manara22}.

Heating in the outer disk becomes dominated by stellar irradiation. In the limit where the disk is optically thick to stellar radiation and vertically isothermal (e.g., nearly optically thin to internal radiation),
\begin{equation}
    T_{\rm irr} \simeq 150 {\rm K} \bigg(\frac{L_\star}{L_{\odot}}\bigg)^{2/7} \bigg(\frac{M_\star}{M_{\odot}}\bigg)^{-1/7} \bigg(\frac{r}{1 \, {\rm au}}\bigg)^{-3/7},
\end{equation}
where $L_\star \propto M_\star^A$ is the stellar luminosity and $A=1.4$--1.9 drawn from MIST models \citep{Dotter16,Choi16} for pre-main sequence stars. We adopt $A = 1.5$ for our study. We can now define the disk temperature:
\begin{equation}
    T_{\rm disk} = {\rm min}(2000 \, {\rm K}, {\rm max} (T_{\rm vis}, T_{\rm irr})) 
\end{equation}
where 2000 K sets the upper limit to the temperature due to the thermostating effect of dust sublimation \citep{Dalessio98}.

\begin{figure}
    \centering
    \includegraphics[width=\linewidth]{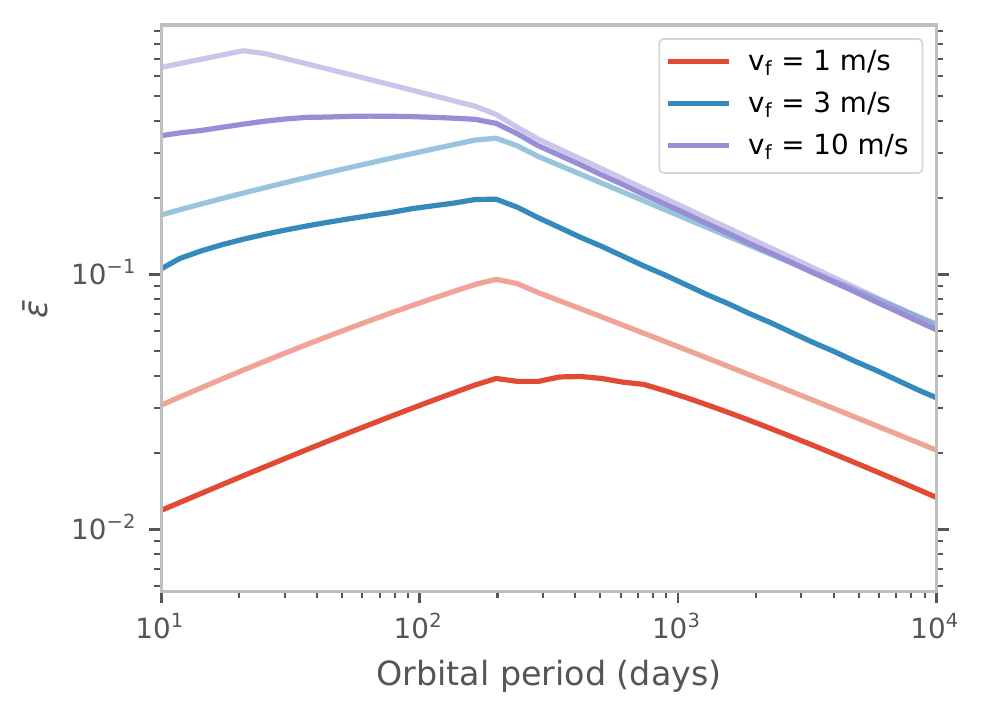}
    \caption{Mass-weighted formation efficiency $\bar{\epsilon}$ vs.~orbital period evaluated for $M_\star=0.5\,M_\odot$, $M_p=1.0\,M_\oplus$, $\alpha_{\rm t}=10^{-3}$, and $\dot{M}_{1\odot}=10^{-8}\,M_\odot\,{\rm yr}^{-1}$. Different colors correspond to varying fragmentation velocity $v_{\rm f}$. The mass-weighted formation efficiency follows the overall behavior of $\epsilon$ evaluated at ${\rm St}_{\rm f}$ (lighter color) with an order unity reduction, i.e. ${\rm d}\bar{\epsilon} / {\rm d}P \sim {\rm d}\epsilon ({\rm St = St_{f}}) / {\rm d}P$. Deviations from this approximation occur for i) $v_{\rm f} = 10\,{\rm m/s}$ at $\sim$20--200 days because accretion transitions from 3D to 2D regime over a spectrum of ${\rm St}$ at different orbital periods and for ii) $v_{\rm f}=1\,{\rm m/s}$ at $\sim$300 days because St$_{\rm f}$ crosses $2\alpha_{\rm t}/\pi$ there such that the grain size distribution is set by only the turbulent regime within $\sim$300 days and by both turbulent and settling regimes beyond it. The kink at $\sim$200 days arises from a change in $\gamma$ from viscous to irradiation-dominated disk heating.}
    \label{fig:avg_eps_per_diag}
\end{figure}

To evaluate $\gamma$, we need to evaluate a pressure profile $P_{\rm g} = \Sigma_{\rm g} c_{\rm s} \Omega_{\rm K}$ where $\Sigma_{\rm g}$ is the disk gas surface density. We adopt the self-similar solution of viscously accreting disk $\Sigma_{\rm g} \propto \nu^{-1} \propto \Omega_{\rm K} \alpha_{\rm t}^{-1} c_{\rm s}^{-2}$ \citep{Lynden-Bell74,Hartmann98} and $P_{\rm g} \propto \Omega_{\rm K}^2 c_{\rm s}^{-1}$. For $T_{\rm disk} \sim T_{\rm vis}$, $P_{\rm g} \propto r^{-51/20}$ so $\gamma = -51/20$. For $T_{\rm disk} \sim T_{\rm irr}$, $P_{\rm g} \propto r^{-39/14}$ so $\gamma = -39/14$. When $T_{\rm disk}$ thermostats, $P_{\rm g} \propto \Omega_{\rm K}^2 \propto r^{-3}$ and so $\gamma = -3$.

With the disk structure in hand, we illustrate in Figure \ref{fig:St_accretion} how the relevant regimes of accretion transition from 3D to 2D and from headwind- to shear-dominated as a function of orbital period, mass of the protoplanet $M_{\rm p}$, and $\alpha_{\rm t}$. For illustrative purpose, we fix $M_\star=0.5M_\odot$ and $\dot{M}_{1\odot}=10^{-8}\,M_\odot\,{\rm yr}^{-1}$. We choose to compare the regimes at the fragmentation limit ${\rm St}_{\rm f}$ as the grain mass distribution is top-heavy in logarithm of St (see equation \ref{eq:avg_eps}). First, we see a general broken power-law in all ${\rm St}$, reflecting a transition from accretion-dominated ($T_{\rm vis}$) to irradiation-dominated ($T_{\rm irr}$) regimes from inner to outer orbits. At fixed $\dot{M}_{1\odot}$, the accretion-dominated regime widens at lower $\alpha_{\rm t}$ because a fixed disk accretion rate with lower turbulence implies larger $\Sigma_{\rm g}$ which in turn raises the disk optical depth and therefore the midplane temperature. Repeating the calculation at higher stellar masses (not shown), we find the accretion-dominated regime widens to longer orbital periods because of the increase in accretion luminosity; irradiation power also increases with stellar mass but not as steeply.

In terms of the mode of pebble accretion, we see that all the different combinations are possible over a plausible parameter space. First, at low $\alpha_{\rm t}$, pebble motions are always dominated by aerodynamic drift rather than the coupling to the viscous gas inflow. As $\alpha_{\rm t}$ increases however, ${\rm St}_{\rm f}$ drops and so the coupling to the gas motion starts to dominate. In general, as the protocore gains mass (and/or at high enough $v_{\rm f}$), pebble accretion becomes more shear-dominated rather than headwind-dominated as the core's gravity starts to dominate the aerodynamic drag. Larger core mass and smaller $\alpha_{\rm t}$ imply larger accretion radius but thinner particle disk scale height and so generally favor 2D accretion. In general, by the time the accretion is two-dimensional, we see that the accretion is shear-dominated with the exception of the lowest $M_{\rm p}$, lowest $\alpha_{\rm t}$ and at long orbital periods (see, e.g., top left panel of Figure \ref{fig:St_accretion}). The range of St, $P$, and $M_{\rm p}$ for which the 2D accretion headwind regime is relevant is independent of $\dot{M}_{1 \odot}$ ($\lesssim  10^{-8} \, M_\odot$ yr$^{-1}$ as the transition from 2D shear to 2D headwind happens in irradiated regions) and although it broadens slightly at lower stellar masses, it still occupies a small part of our parameter space.

\begin{figure*}
    \centering
    \includegraphics[width=\textwidth]{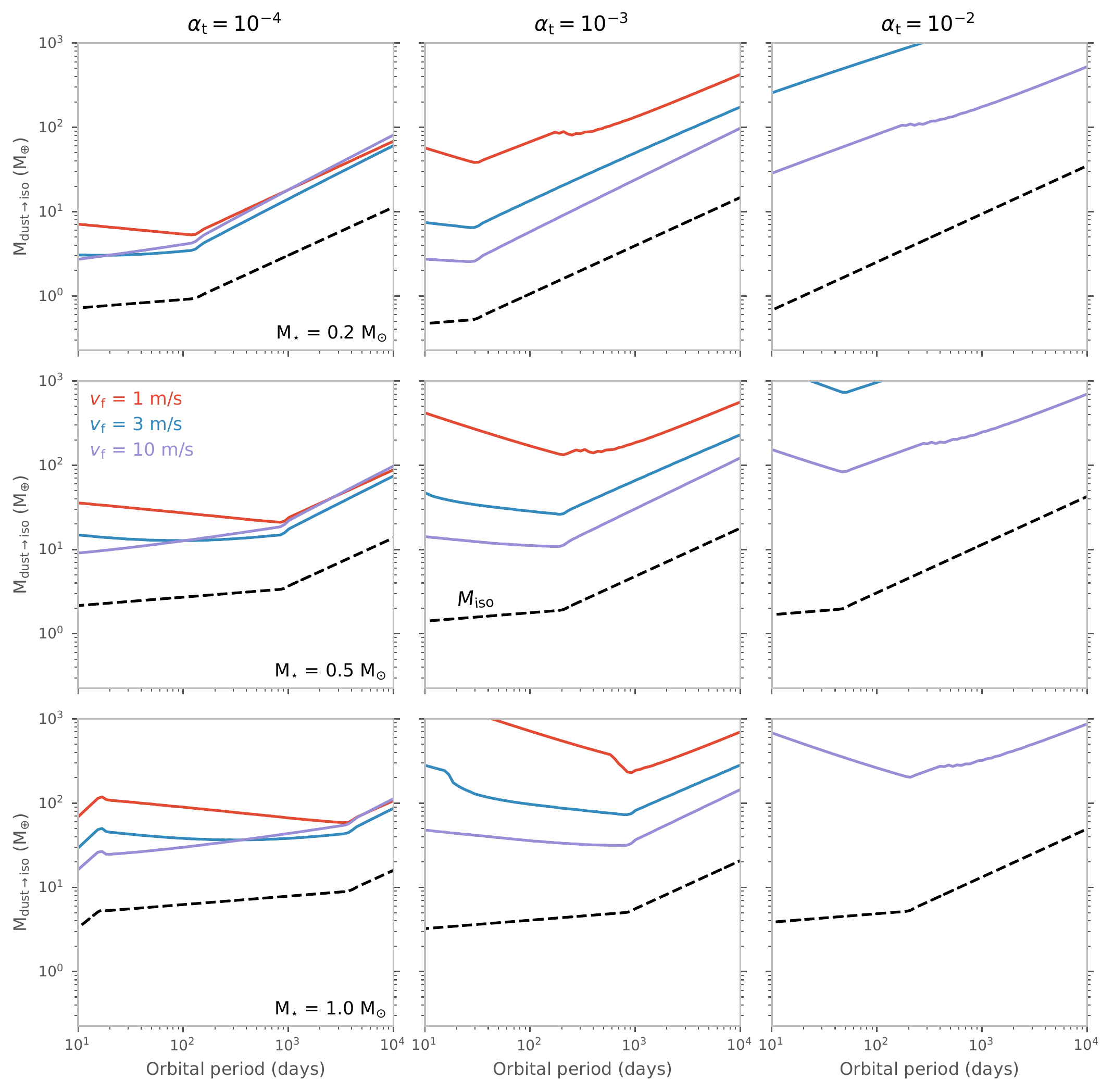}
    \caption{Required dust mass to reach pebble isolation mass $M_{\rm iso}$ (dashed lines) at each orbital period for a variety of $\alpha_{\rm t}$ (different columns), $v_{\rm f}$ (different colors) and $M_\star$ (different rows). The accretion rate is fixed at $\dot{M}_{1\odot}=10^{-8}M_\odot\,{\rm yr}^{-1}$. In general, more mass is required at higher $\alpha_{\rm t}$, lower $v_{\rm f}$, and higher $M_\star$, reflecting lower accretion efficiencies. In the inner disk that is viscously-heated, $M_{\rm dust \rightarrow iso}$ usually drops with orbital period whereas in the outer disk that is irradiation-heated, $M_{\rm dust \rightarrow iso}$ tends to increase with orbital period. Exceptions are discussed in more detail in the main text.}
    \label{fig:Mdust_Mst_Per}
\end{figure*}

\section{Results}
\label{sec:results}

\subsection{Expected Behaviors of Required Dust Mass}
\label{sec:analy_expect}
Before we present our numerical results, we derive how $\bar{\epsilon}$ is expected to scale with $M_\star$ and orbital period $P$. We start with $T_{\rm disk}$:
\begin{equation}
    T_{\rm vis} \propto M_\star^{2 B / 5}  P^{-3/5} \dot{M}_{1 \odot}^{2/5} \alpha_{\rm t}^{-1/5} 
    \label{eq:T_vis_scaling}
\end{equation}
\begin{equation}
    T_{\rm irr} \propto M_\star^{2 (A - 1) / 7} P^{-2/7}.
    \label{eq:T_irr_scaling}
\end{equation}

The 2D headwind regime is applicable only for a small range of St and $M_{\rm p}$ (see Figure \ref{fig:St_accretion}) and most of the mass is acquired in 2D shear and 3D accretion regimes. By the time the protocore reaches 2D shear accretion, ${\rm St}$ is large enough for radial drift to be dominated by aerodynamic drag and so from equation \ref{eq:acc-eff},
\begin{equation}
    \epsilon_{\rm 2D, sh} \propto \gamma^{-1} \bigg(\dfrac{M_{\rm p}}{M_\star}\bigg)^{2/3} \bigg(\dfrac{v_{\rm K}}{c_{\rm s}}\bigg)^{2} {\rm St}^{-1/3}.
    \label{eq:eps_2d_shear_scale}
\end{equation}
Since $2|\gamma|/3 > 1$, there are three possibilities for 3D accretion:
\begin{align}
    \epsilon_{\rm 3D} &\propto
    \begin{cases}
        \gamma^{-1} \bigg(\dfrac{M_{\rm p}}{M_\star}\bigg) \bigg(\dfrac{v_{\rm K}}{c_{\rm s}}\bigg)^{3} \bigg(\dfrac{\rm St}{\alpha_{\rm t}}\bigg)^{1/2} & {\rm St} > \alpha_{\rm t} \\
        \gamma^{-1} \bigg(\dfrac{M_{\rm p}}{M_\star}\bigg) \bigg(\dfrac{v_{\rm K}}{c_{\rm s}}\bigg)^{3} & \dfrac{3}{2} \dfrac{\alpha_{\rm t}}{|\gamma|} < {\rm St} < \alpha_{\rm t} \\
        \bigg(\dfrac{M_{\rm p}}{M_\star}\bigg) \bigg(\dfrac{v_{\rm K}}{c_{\rm s}}\bigg)^{3} \dfrac{\rm St}{\alpha_{\rm t}} & {\rm St} < \dfrac{3}{2} \dfrac{\alpha_{\rm t}}{|\gamma|}
    \end{cases}
    \label{eq:eps_3d_scale}
\end{align}

An examination of Equations~\ref{eq:eps_2d_shear_scale} and \ref{eq:eps_3d_scale} at fixed orbital period $P$ shows that accretion efficiency primarily depends on $M_\star$ via the disk aspect ratio $c_{\rm s} / v_{\rm K} = H_{\rm g} / r$. At a fixed $P$, $c_{\rm s} / v_{\rm K}$ is lower around lower mass stars, which implies a higher pebble accretion rate (pebble disk is less puffy) and a pebble drift rate $\dot{M}_{\rm drift}$ that is either slower for lower $M_\star$ (St $< 3 \alpha_{\rm t} / 2 |\gamma|$) or independent of $M_\star$ (St $> 3 \alpha_{\rm t} / 2 |\gamma|$). Thus, the accretion efficiency is boosted for lower $M_\star$ as a result of lower $c_{\rm s} / v_{\rm K}$.

The second condition on $\epsilon_{\rm 3D,vis}$ in Equation~\ref{eq:eps_3d_scale} only occurs for a narrow range of St and so is not dominant. As a result, we will not consider it for subsequent analytical derivations. Since $\bar{\epsilon}$ is more weighted towards larger grains that contain most of the mass, we now substitute St = St$_{\rm f}$ solely to derive the scaling relationship of $\bar{\epsilon}$ with respect to physical parameters. As illustrated in Figure \ref{fig:avg_eps_per_diag}, this is an adequate substitution to express the scaling relationships (not the absolute normalization) except near the transition between 3D and 2D accretion---because each ${\rm St}$ makes its transition at a spectrum of orbital periods---and where St$_{\rm f}$ crosses the turbulent vs.~settling regime. We note that we use the full range of St in our numerical calculations.

Substituting St with St$_{\rm f}$, using the temperature scaling for viscously-heated regions of the disk (Equation~\ref{eq:T_vis_scaling}, relevant for inner orbits), and writing $v_{\rm K}$ in terms of $M_\star$ and $P$ in Equations~\ref{eq:eps_2d_shear_scale} and \ref{eq:eps_3d_scale}, we find that
\begin{equation}
    \epsilon_{\rm 2D, vis} \propto M_{\rm p}^{2/3}M_\star^{-4B/15}P^{-4/15}\dot{M}_{1\odot}^{-4/15}v_{\rm f}^{-2/3}\alpha_{\rm t}^{7/15},
    \label{eq:eff_2d_vis_scl}
\end{equation}
and
\begin{align}
    &\epsilon_{\rm 3D, vis} \propto \nonumber \\
    &\begin{cases}
        M_{\rm p}M_\star^{-4B/5}P^{1/5}\dot{M}_{1\odot}^{-4/5}v_{\rm f}\alpha_{\rm t}^{-3/5} & {\rm St_{f}} > \alpha_{\rm t}  \\
        M_{\rm p}M_\star^{-B}P^{1/2}\dot{M}_{1\odot}^{-1}v_{\rm f}^2\alpha_{\rm t}^{-3/2} & {\rm St_{f}} < \dfrac{3}{2} \dfrac{\alpha_{\rm t}}{|\gamma|}
    \end{cases}
    \label{eq:eff_3d_vis_scaling}
\end{align}
The scaling of $\epsilon$ with respect to $P$ flips sign between 2D and 3D accretion because this dependence on the orbital period is primarily set by St rather than the disk aspect ratio in viscously heated regions (St$_{\rm f} \propto P^{3/5}$ compared to $v_{\rm K} / c_{\rm s} \propto P^{-1/30}$). In the 2D accretion regime, larger St$_{\rm f}$ in the outer orbit implies a vigorous radial drift of pebbles, which makes it hard for the protocore to capture them. Under 3D accretion, larger St$_{\rm f}$ boosts $\dot{M}_{\rm peb}$ more so than $\dot{M}_{\rm drift}$ so that $\epsilon_{\rm 3D,vis}$ rises with $P$. In both cases, $\epsilon$ drops with $M_\star$ for all plausible range of $B=1.5$--3.1 because both $M_{\rm p}/M_\star$ and $v_{\rm K}/c_{\rm s}$ decrease with $M_\star$. In particular, note the stronger than inversely linear dependence of $\epsilon_{\rm 3D,vis}$ on $M_\star$.

Making similar substitutions in Equations~\ref{eq:eps_2d_shear_scale} and \ref{eq:eps_3d_scale} but using the temperature scaling for irradiation dominated regions of the disk (Equation~\ref{eq:T_irr_scaling}), we obtain
\begin{equation}
    \epsilon_{\rm 2D,irr} \propto M_{\rm p}^{2/3}M_\star^{-4(A-1)/21}P^{-10/21}v_{\rm f}^{-2/3}\alpha_{\rm t}^{1/3}, 
    \label{eq:eff_2d_irr_scl}
\end{equation}
and
\begin{equation}
    \epsilon_{\rm 3D, irr} \propto
    \begin{cases}
        M_{\rm p}M_\star^{-4(A - 1)/7}P^{-3/7}v_{\rm f}\alpha_{\rm t}^{-1} & {\rm St_{f}} > \alpha_{\rm t} \\
        M_{\rm p}M_\star^{-5(A-1)/7}P^{-2/7}v_{\rm f}^2\alpha_{\rm t}^{-2} & {\rm St_{f}} < \dfrac{3}{2} \dfrac{\alpha_{\rm t}}{|\gamma|}
    \end{cases}
    \label{eq:eff_3d_irr_scaling}
\end{equation}
Both $\epsilon_{\rm 2D,irr}$ and $\epsilon_{\rm 3D, irr}$ now fall off with $P$ because $v_{\rm K}/c_{\rm s}$ declines more rapidly with $P$ in irradiated regions ($v_{\rm K}/c_{\rm s} \propto P^{-4/21}$ in the irradiation region compared to $v_{\rm K}/c_{\rm s} \propto P^{-1/30}$ in the viscous region). We also see the drop with $M_\star$, albeit more weakly compared to viscously-heated region for the entire range of $A=1.4$--1.9, due to the weaker dependence of $T_{\rm irr}$ on $M_\star$. 

While $M_{\rm iso}$ (and potentially $M_0$) also depends on $M_\star$ and $P$, we expect its contribution to $M_{\rm disk \rightarrow iso}$ to be minimal compared to $\epsilon$ since equation~\ref{eq:mdust-to-iso} implies $M_{\rm dust \rightarrow iso} \propto {\rm ln} \, (M_{\rm iso} / M_0)$ in the 3D regime and $M_{\rm dust \rightarrow iso} \propto (M_{\rm iso}^{1/3} - M_0^{1/3})$ in the 2D regime, both of which are weak dependencies.

\begin{figure*}[t]
    \centering
    \includegraphics[width=0.32\linewidth]{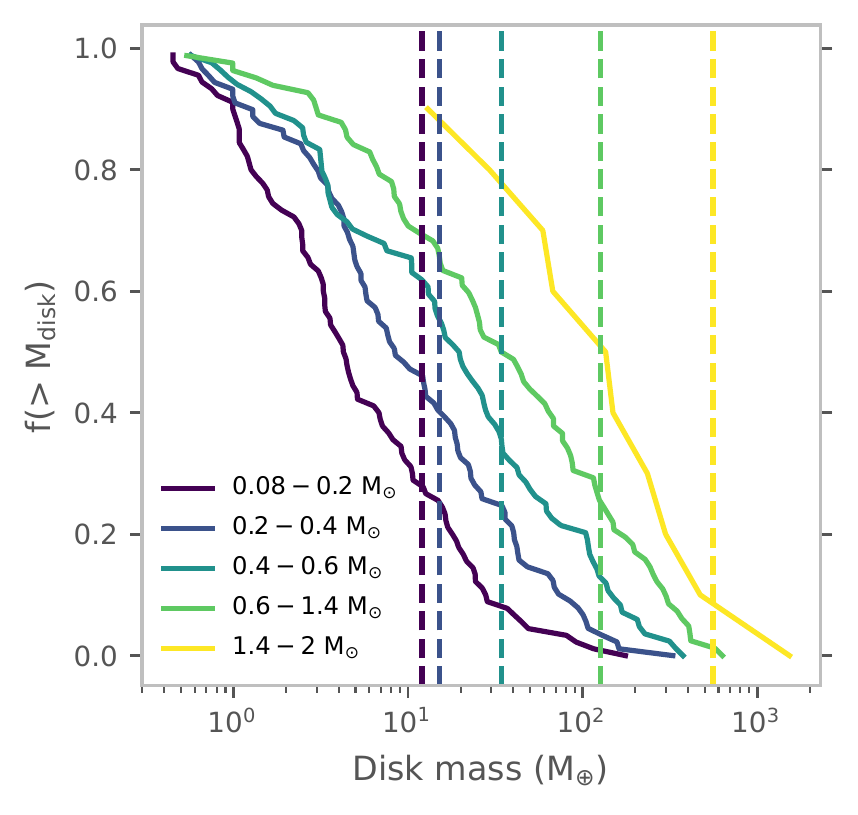}
    \includegraphics[width=0.33\linewidth]{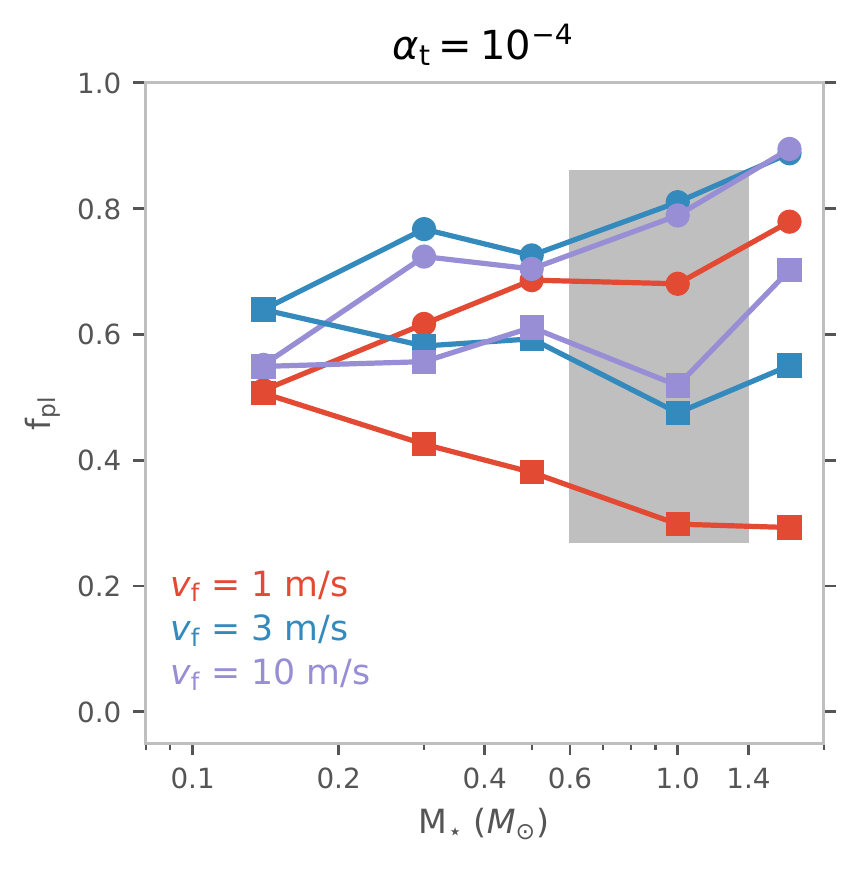}
    \includegraphics[width=0.33\linewidth]{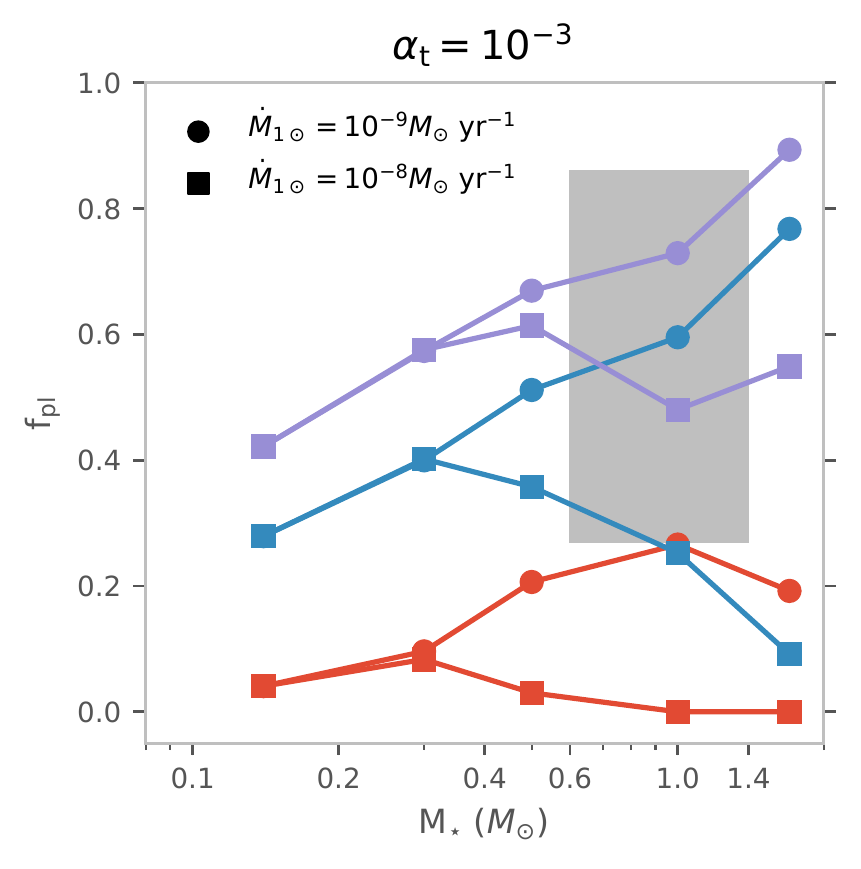}
    \caption{Left: the cumulative distribution function (CDF) of disk masses from \citet{Manara22} augmented by factors of $\sim$3 to have the maximum disk mass match that of Class I disks measured in \citet{Tobin20}, color-coded with respect to stellar mass bins. The vertical dashed lines represent $M_{\rm dust \rightarrow iso}$ in the inner orbits (taken as the larger mass between 30 and 100 days) evaluated for $\alpha_{\rm t}=10^{-3}$, $\dot{M}_{1\odot}=10^{-8}\,M_\odot\,{\rm yr}^{-1}$, and $v_{\rm f}=$3 m/s, at the midpoint stellar mass for each given bin.
    Middle: the fraction of disks with enough mass to create an isolation mass in the inner orbits (taken as the larger mass between 30 and 100 days) for $\alpha_{\rm t}=10^{-4}$. The vertical gray rectangle shows the range of measured fraction of FGK stars with {\it Kepler} planets from \citet{Zhu18} and \citet{Yang20}. Right: same as the middle panel but for $\alpha_{\rm t}=10^{-3}$. 
    All models but one produce $f_{\rm pl}$ matching the measured value for FGK stars. In general, the fraction of disks that can create inner cores rises as the stellar mass decreases from $\sim$1$M_\odot$ to $\sim 0.3 - 0.5 M_\odot$ if the cores form in viscously heated regions in the 3D regime (e.g., $\dot{M}_{1,\odot}=10^{-8} \, M_\odot \, {\rm yr}^{-1}$).
    For $\dot{M}_{1,\odot}=10^{-9} \, M_\odot \, {\rm yr}^{-1}$, planetary cores form in irradiated regions of the disk at $P = 100$ days for $\alpha_{\rm t} = 10^{-3}$ and 2D accretion becomes important for $\alpha_{\rm t} = 10^{-4}$; $\epsilon$ has a shallower dependence on $M_{\star}$ in these conditions.}
    \label{fig:fdisk_fpl}
\end{figure*}

Figure \ref{fig:Mdust_Mst_Per} illustrates the overall behavior of $M_{\rm dust \rightarrow iso}$ as a function of orbital period over a range of $\alpha_{\rm t}$, $M_\star$, and $v_{\rm f}$. Here, $M_{\rm dust \rightarrow iso}$ is calculated using the range of grain sizes present in fragmentation-limited regions of protoplanetary disks (St $\in [10^{-6}, {\rm St}_{\rm f}]$, see discussion around Equations~\ref{eq:avg_eps}-\ref{eq:mdust-to-iso}). Higher $\epsilon$ implies lower required dust mass to create $M_{\rm iso}$ at a given orbital period so if $\epsilon$ decreases with $P$, the corresponding $M_{\rm dust \rightarrow iso}$ rises with $P$. In general, we see rising $M_{\rm dust \rightarrow iso}$ with $P$ in the outer orbit where disk heating is dominated by stellar irradiation (c.f.~equation \ref{eq:eff_2d_irr_scl}) and falling $M_{\rm dust \rightarrow iso}$ with $P$ in the inner orbit where the disk is viscously heated (c.f.~equation \ref{eq:eff_3d_vis_scaling}). Exception arises at high $v_{\rm f}$ and $\alpha_{\rm t} = 10^{-4}$ in the inner orbit which is when the accretion turns to 2D (c.f.~equation \ref{eq:eff_2d_vis_scl}). The dips in $M_{\rm dust \rightarrow iso}$ (for $P$ in the range $\sim$200-1000 days, $v_{\rm f}=1$~m/s, $\alpha_{\rm t}=10^{-3}$, middle column in Figure~\ref{fig:Mdust_Mst_Per} and $v_{\rm f}=10$ m/s, $\alpha_{\rm t}=10^{-2}$, right column in Figure~\ref{fig:Mdust_Mst_Per}; for $P \sim 20$ days, $v_{\rm f}=3$ m/s, $\alpha_{\rm t}=10^{-3}$, middle row, bottom panel of Figure~\ref{fig:Mdust_Mst_Per}) manifest due to both turbulent and settling regimes of St coming into effect (c.f.~Figure \ref{fig:avg_eps_per_diag}). Overall the required dust mass is larger for particles that more easily fragment (lower $v_{\rm f}$) since this limits the maximum St and pebble accretion is slower for smaller St. However, we see an exception at $\alpha_{\rm t}=10^{-4}$ where the required dust mass becomes comparable and slightly larger for $v_{\rm f}=10$m/s compared to 3 m/s in the outer orbit as they are both under 2D accretion which is less efficient at higher St due to strong radial drift past the protocore (c.f.~equation \ref{eq:acc-eff}).

\subsection{Fraction of Disks that Can Create Inner Planets}

Even though the pebble accretion efficiency is higher for lower mass stars in all the different regimes we study (Equations~\ref{eq:eff_2d_vis_scl}$-$\ref{eq:eff_3d_irr_scaling}), the feasibility of forming planetary cores depends on the available pebble mass as well, which generally decreases around lower mass stars. We therefore compare our $M_{\rm dust \rightarrow iso}$ against the distribution of measured disk masses to compute the fraction of disks that have enough mass to create inner cores $f_{\rm pl}$, which we directly translate into the fraction of stars that have {\it Kepler} planets \citep[e.g.,][]{Zhu18_occur,Yang20}, a metric that is distinct from the commonly quoted planet occurrence rate defined as the number of planets per star. 

Using the measured disk masses tabulated by \citet{Manara22}, we build a cumulative distribution function (CDF) defined as the fraction of disks with masses greater than a given $M_{\rm disk}$ (Figure~\ref{fig:fdisk_fpl}, left panel). The CDFs are divided into a range of stellar masses with the lower and upper limits of each interval set to (0.08, 0.20, 0.40, 0.60, 1.40, 2.00) $M_\odot$, chosen to represent different stellar spectral types. These are evolved disks ($\gtrsim$1 Myr) but the formation of planetary cores could very well begin earlier \citep[e.g.,][]{Najita14, Manara18} and so we multiply all disk masses by a factor of 3 so that the $3\sigma$ (99.7th percentile) value of disk mass in our sample matches the $3\sigma$ value of Class I disk masses measured by VLA in \cite{Tobin20}. A factor of 3 increase in dust mass between Class II and Class I disks is also in agreement with existing results from comparison of different disk classes in various star forming regions \citep[e.g.,][]{Tychoniec2020}. See Appendix~A for more details and the effects of our choices on the disk mass CDFs and $f_{\rm pl}$.

From Figure \ref{fig:Mdust_Mst_Per}, we first note that at $\alpha_{\rm t}=10^{-2}$, the required dust mass is often $\gtrsim 10^3\,M_\oplus$, which is beyond the augmented masses of our disk sample and so we rule out $\alpha_{\rm t}=10^{-2}$. For the remaining runs, we compute $f_{\rm pl}$ for each stellar mass (taken as the midpoint of our stellar mass intervals) by plugging in the corresponding $M_{\rm dust \rightarrow iso}$ into the inverse function of the CDF. Given that the pebble accretion efficiency is generally 1--10\% (see Figure \ref{fig:Mdust_Mst_Per}), planetary cores at each location act as inefficient filters of pebbles allowing for the growth of multiple cores. While quantifying the exact multiplicity of cores is beyond the scope of this paper, within our framework, we can state that the region of planet formation is dictated by where the total disk dust mass is above $M_{\rm dust,iso}$ and so a system can potentially nucleate more cores if it contains larger initial dust mass reservoir. We therefore take the larger of the $M_{\rm dust \rightarrow iso}$ values corresponding to $P = 30$ and 100 days in order to capture the dust mass required to populate the inner disk with multiple planetary cores at least down to 30 days. Our qualitative results are not sensitive to the exact choice of this period range. The middle and the right panels of Figure \ref{fig:fdisk_fpl} show the resulting $f_{\rm pl}$ vs.~$M_\star$ for $\alpha_{\rm t} = 10^{-4}$ and $10^{-3}$ respectively. At $\dot{M}_{1\odot} = 10^{-8} \,M_\odot \,{\rm yr}^{-1}$, the fraction of disks that can create inner cores generally rises as stellar mass decreases from $\sim 1 M_\odot$ to $\sim 0.3 - 0.5M_\odot$,\footnote{The rise in $f_{\rm pl}$ beyond $1 M_\odot$ is likely because of the poorly characterized disk mass CDF for massive stars due to limited sample size. The number of protostars in the different stellar mass bins are 90 ($0.08-0.2 M_\odot$), 89 ($0.2-0.4 M_\odot$), 84 ($0.4-0.6 M_\odot$), 82 ($0.6-1.4 M_\odot$), and 10 ($1.4-2 M_\odot$).)} reflecting the increase in $\epsilon$ with smaller $M_\star$ for pebble accretion in viscously heated regions in the 3D regime (Equation~\ref{eq:eff_3d_vis_scaling}).

The trend tends to reverse at lower $M_\star$ because around these lower mass stars, even the inner orbits become irradiation-dominated and $\epsilon_{\rm 3D,irr}$ is much more weakly dependent on $M_\star$ (Equation~\ref{eq:eff_3d_irr_scaling}). The corresponding $M_{\rm dust \rightarrow iso}$ becomes near constant and since the disk mass distribution around lower $M_\star$ is more bottom-heavy, $f_{\rm pl}$ decreases. We do not see this overturn in $f_{\rm pl}$ at lower stellar masses when the viscous heating dominates at the relevant orbital periods (e.g., $\alpha_{\rm t} = 10^{-4}$ and $\dot{M}_{1\odot} = 10^{-8} \,M_\odot \,{\rm yr}^{-1}$). At $\dot{M}_{1\odot}=10^{-9} \, M_\odot \, {\rm yr}^{-1}$, $f_{\rm pl}$ often rises with $M_\star$, because of the weak stellar mass dependence of $\epsilon$ in the 2D accretion regime ($\alpha_{\rm t} = 10^{-4}$, Equation~\ref{eq:eff_2d_vis_scl}) or in the irradiated region ($\alpha_{\rm t} = 10^{-3}$, Equation~\ref{eq:eff_3d_irr_scaling}); the resulting $M_{\rm dust \rightarrow iso}$ is too similar compared to the difference in the disk mass CDFs, and thus $f_{\rm pl}$ rises with stellar mass.

We compare our $f_{\rm pl}$ with the the measured fraction of FGK stars with {\it Kepler} planets which ranges from $\sim$30$\pm$3\% \citep{Zhu18_occur} to $\sim$73$\pm$13\% \citep{Yang20} where the difference arises from different samples of host stars and different models for correcting for intrinsic planet multiplicity and the correlation in periods and radii of adjacent planet pairs \citep[see][their Section 5.1]{Yang20}. Under our scheme, all but one model can reproduce the fraction of stars harboring close-in {\it Kepler} planets (see Figure \ref{fig:fdisk_fpl}).

\subsection{Connection with Outer Planets}

\begin{figure}
    \centering
    \includegraphics[width=0.9\linewidth]{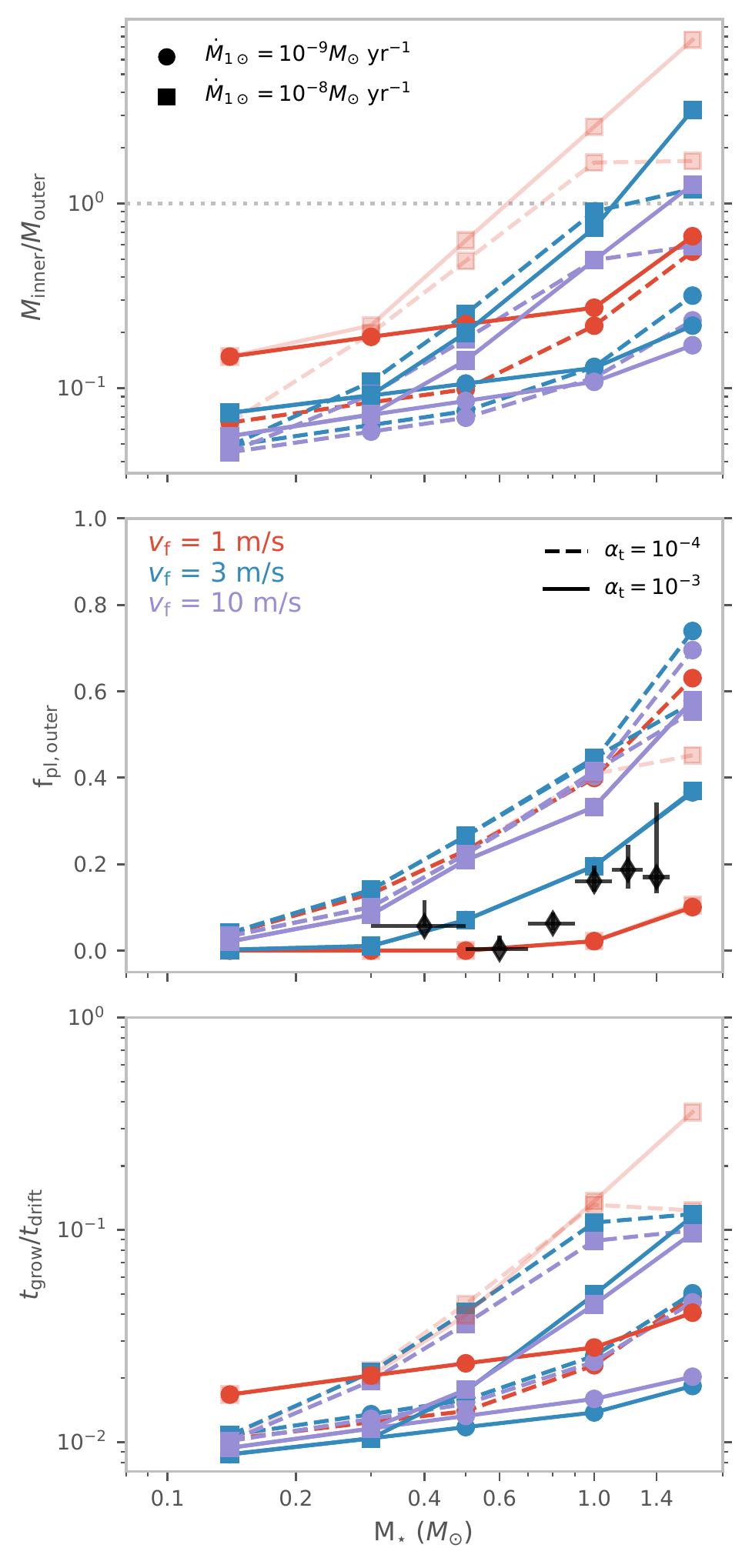}
    \caption{Top: the ratio of required dust mass to create the inner cores to the required dust mass to create an outer isolation mass at 5 AU as a function of stellar mass. Middle: the fraction of disks with enough mass to create an isolation mass at 5 AU $f_{\rm pl,outer}$ vs.~stellar mass. Bottom: the ratio of growth timescale of inner cores to the drain-out timescale of the dust mass that was drifted in before the creation of an 5 AU isolation mass. 
    In diamonds, we plot the occurrence rate of giants (100--6000$M_\oplus$) between 1-5 AU computed by \citet{Fulton21} with the errorbar delimiting 15.9--84.1\% confidence intervals, reproduced from their Figure 7 (see footnote 3). In general, disks around higher mass stars are more likely to nucleate an outer planet and when they do, more than enough mass can drift in to the inner disk to create short-period planetary cores for $M_\star \leq M_\odot$ (exception shown in light color), the latter of which can assemble before all the solids drain out to the inner edge of the disk.}
    \label{fig:Jup_cnxn}
\end{figure}

We now compute $M_{\rm dust \rightarrow iso}$ and the fraction of disks that can create an isolation mass at 5 AU $f_{\rm pl,outer}$. For our parameters, this isolation mass is always $\gtrsim 9$--10$M_\oplus$ which is large enough to nucleate a gas giant at this distance, with the exact runaway mass subject to local opacity and metal content \citep[e.g.,][]{Piso15, Lee15, Venturini15, Chachan21}. First, we verify that the amount of dust mass required to create this outer core ($M_{\rm outer}$) is larger than that required to create inner cores ($M_{\rm inner}$, top panel of Figure~\ref{fig:Jup_cnxn}). Since $M_{\rm dust \rightarrow iso}$ rises with orbital period in irradiated regions (Figure~\ref{fig:Mdust_Mst_Per}), this condition is met by all our models for $M_\star < M_\odot$ and by most of them for $M_\star \geq M_\odot$. When $M_{\rm inner} < M_{\rm outer}$, a disk that has enough dust mass ($\geq M_{\rm outer}$) to create outer giant planet cores will invariably create both inner and outer planets. This is in agreement with the previous results of \citet{Lin18} and \citet{Chachan22}. The two models for which this is not true for Sun-like stars correspond to pebble accretion happening in the 3D regime in a viscously heated region that is large in extent or for ${\rm St_{f}} < 3 \alpha_{\rm t} / 2 |\gamma|$ ($\alpha_{\rm t} = 10^{-4}$ \& $10^{-3}$, $\dot{M}_{1\odot} = 10^{-8} \,M_\odot \,{\rm yr}^{-1}$, $v_{\rm f} = 1$~m/s; light shade in Figure \ref{fig:Jup_cnxn}). We note that the formation of outer giant cores at the measured occurrence rate is only feasible for the former of these two models.

Furthermore, we see that $f_{\rm pl,outer}$ rises with $M_\star$ (middle panel of Figure \ref{fig:Jup_cnxn}) in agreement with the observed trends \citep[e.g.,][]{Fulton21}.\footnote{We note that our calculation provides the fraction of disks that can form giants whereas \citet{Fulton21} measures the occurrence rate of giants. While observational measurements of the former are not yet available, the average cold giant multiplicity is estimated to be $1.27 \pm 0.12$ \citep{Zhu2022}, suggesting that average systems are expected to harbor single cold giant. It is therefore reasonable to assume that the cold giant planet occurrence rate is not too different from the fraction of stars with cold giant planets.} In almost all cases, the accretion is in the 2D, irradiation-dominated regime and although $\epsilon_{\rm 2D,irr}$ falls with $M_\star$ (Equation~\ref{eq:eff_2d_irr_scl}), it is a weak dependence and the distribution of $M_{\rm disk}$ is more top-heavy around more massive stars so that $f_{\rm pl,outer}$ will be larger. Compared to the measured occurrence rate of outer gas giants from \citet{Fulton21}, our $f_{\rm pl,outer}$ tends to be large. We note that we are simply calculating the fraction of disks with enough mass to create an isolation mass by pebble accretion. Accounting for the time required for the completion of pebble accretion and subsequent gas accretion to form planets $\gtrsim 100 M_{\oplus}$ would likely reduce our $f_{\rm pl,outer}$ given finite disk lifetimes \citep{Mamajek2009, Michel21} and varying metal content (opacity), thus bringing it into better agreement with the measured occurrence rate of cold Jupiters.

We deduce the effect of filtering by an outer planet proposed by \citet{Mulders21} by considering the inner disk to be decoupled from the outer disk once the outer planet reaches its isolation mass. Calculating the growth timescale $t_{\rm grow}$ of an inner core ($M_{\rm iso}/\dot{M}_{\rm peb}$) out of the solids that were drifted in ($M_{\rm solids}$=$M_{\rm dust \rightarrow iso}$(5 AU) - $M_{\rm iso}$(5 AU)) and comparing $t_{\rm grow}$ to the drain out timescale of $M_{\rm solids}$ ($t_{\rm drain} =  r/v_{\rm r}$) at $P = 100$ days ($\sim$ the outer period at which current surveys are sensitive to super-Earths), we find that $t_{\rm grow} < t_{\rm drain}$ in all the cases we consider (see bottom panel of Figure \ref{fig:Jup_cnxn}), suggesting that the inner cores would be formed before all the solids disappear. We verify that filtering is only relevant when $v_{\rm f}$ is smaller at short orbital periods, and so our result is consistent with the ``No Snow Line'' curve in Figure 1 of \citealt{Mulders21}. While the silicate grains within the iceline are often considered to have lower $v_{\rm f}$ than the ice-coated grains in the outer orbit \citep[e.g.,][]{Pinilla16, Drazkowska17}, recent laboratory and theoretical studies suggest $v_{\rm f}$ is more uniform \citep[e.g.,][]{Musiolik19, Kimura20}.

Considering that both $M_{\rm inner} / M_{\rm outer}$ and $t_{\rm grow} / t_{\rm drain}$ (top and bottom panels of Figure~\ref{fig:Jup_cnxn}) are lower around lower mass stars, our results suggest that it is easier to form super-Earths interior to cold-giants around lower mass stars. Our calculation therefore predicts that the correlation between the inner super-Earths and outer giants that is observed for FGK stars persists and perhaps even strengthens for M dwarfs.

\section{Discussion and Conclusion}
\label{sec:discussion}

By combining the theory of pebble accretion with the measured disk masses, we have demonstrated that the fraction of disks that can create inner planetary cores rises towards lower mass stars down to $\sim 0.3 - 0.5M_\odot$ over a plausible range of $v_{\rm f}$ and $\alpha_{\rm t}$ for a typical disk accretion rate of $\dot{M}_\star = 10^{-8}\,M_\odot\,{\rm yr}^{-1}(M_\star/M_\odot)^2$, in agreement with the observed enhancement in the occurrence rate of inner super-Earths around M dwarfs compared to FGK dwarfs \citep[e.g.,][]{Dressing15,Mulders15,Hsu20}. Since core formation by pebble accretion likely happens during the earlier stages of a disk's lifetime, $\dot{M}_{1 \odot} \sim 10^{-8}\,M_\odot\,{\rm yr}^{-1}$ is a reasonable choice for the accretion rate of a solar mass star. We further showed that disks that have enough mass to create an isolation mass at 5 AU will be able to drift in more than enough solids to the inner disk to create small planets there, explaining the observed correlation between inner super-Earths and outer giants \citep[e.g.,][]{Zhu18, Bryan19}. Unlike the inner planets, the fraction of disks that can create a large outer planet drops towards lower mass stars, in agreement with the observed reduction of cold Jupiter occurrence rate around cooler stars \citep[e.g.,][]{Clanton14, Fulton21}. Our framework therefore provides a unifying solution that can reconcile all three observations, with a caveat that it is sensitive to the measurements of disk masses. More accurate and complete measurements of disk properties would be welcome to further verify and refine our calculations.

In general, we predict the fraction of disks that can create inner planets up to their isolation masses $\gtrsim$1--2$M_\oplus$ to drop around stars $M_\star < 0.3$--0.5$M_\odot$; it remains possible to create smaller cores. Since this is about the mass below which short-period planets would emerge rocky \citep{Lee21, Lee22}, our model expects a fall in the occurrence rate of mini-Neptunes around low mass stars but not necessarily that of terrestrial planets, in line with recent observational evidence \citep[e.g.,][]{Brady22, Ment23}. A caveat to this statement is that we have limited our analysis to pebble accretion which sets the initial mass of planetary cores. A more quantitative verification would require bridging this early stage of core formation to the final mass doubling effected by giant impacts in the late stage of disk evolution \citep[e.g.,][]{Lee16, Dawson16} since how much gas a planet ends up with is primarily determined by its final core mass \citep[e.g.,][]{Lee19}.

If we consider a warmer giant ($\lesssim$3 AU), our framework would expect a weakening of the inner and outer planet correlation as $M_{\rm dust \rightarrow iso}$ becomes similar between $\sim$100 days and $\lesssim$3 AU (and in some cases, the required dust mass to create the inner core can exceed), in qualitative agreement with \citet{Rosenthal22}. A distinct prediction of our calculation is that, as long as colder Juipters exist, their correlation with the inner super-Earths remains and potentially enhances even around M dwarfs.

We close with a comment on the effect of multiple outer giants. In our analysis, we have focused on a single giant, which is consistent with the estimated low average cold giant multiplicity 1.27$\pm$0.12 \citep{Zhu2022}. The same study reports 8 systems with 2 cold giants; furthermore, our solar system harbors two cold {\it gas} giants (Jupiter and Saturn). What is the expected effect of multiple cold giants in the context of our study? Given that the pebble accretion efficiency at the location of these giants is $\sim$10\%, if $M_{\rm inner}/M_{\rm outer} \gtrsim 0.8$, having two giants can inhibit the assembly of isolation mass cores in the inner orbits, at least by pebble accretion. Qualitatively, this is consistent with the isotopic constraints suggesting Earth and Mars are mostly formed out of non-carbonaceous material (i.e., inner disk material) with only a trace amount of carbonaceous chondrites, which would have had to drift in from the outer orbits \citep[e.g.,][]{Warren11,Burkhardt21}. With pebble accretion minimized, the overall growth of planetary cores would also likely slow down, which may help explain why our solar system ended up with only small terrestrial planets rather than super-Earths/mini-Neptunes. In general, depending on $M_{\rm inner}/M_{\rm outer}$, multiple giants can significantly reduce the amount of solid that drift into the inner disk and limit the growth of planetary cores there. So under the framework presented here, we would expect a weakening correlation between {\it multiple} cold Jupiters and inner super-Earths/mini-Neptunes. Verifying this preliminary hypothesis by quantifying the effect of cold giant multiplicity is a subject of future work.

\section*{acknowledgments}
We thank the referee for a speedy and careful report that helped improve our paper. We also thank Madison Brady, Marta Bryan, Eugene Chiang, Paul Dalba, Rebekah Dawson, Beibei Liu, Max Moe, and Gijs Mulders for useful discussions and giving comments on an earlier version of this paper. E.J.L thanks Joanna Dr{\k{a}}{\.z}kowska and Francis Nimmo for useful exchanges on pebble trapping and solar system formation at the 2023 Gordon Research Conference. 

Y.C. acknowledges support from the Natural Sciences and Engineering Research Council of Canada (NSERC) through the CITA National Fellowship and the Trottier Space Institute through the TSI Fellowship. E.J.L. gratefully acknowledges support by NSERC, by le Fonds de recherche du Québec – Nature et technologies (FRQNT), by the Trottier Space Institute, and by the William Dawson Scholarship from McGill University. 

\appendix
\counterwithin{figure}{section}

\section{Disk mass CDF}
Since we need both disk dust mass and stellar mass for our study, we use Class II disk dust masses compiled by \cite{Manara22} from measurements of disk masses in different star forming regions from the literature. These disk masses are estimated from ALMA observations assuming that the emission is optically thin, the dust opacity is given by $2.3 (\nu / 230$ GHz) cm$^2$/g, and a single dust temperature of 20 K. We refer the reader to \cite{Manara22} for more details. We limit our sample to systems that have both measured disk masses (excluding upper limits) and stellar masses, which gives a sample size of 402.

Since Class II disks are known to already have undergone significant dust evolution and to contain too little mass to form the observed planetary systems \citep{Najita14}, we must look to disks at earlier stages in their evolution to obtain a better estimate of the initial dust mass. To do so, we turn to disk dust masses measured for Class I disks in the Orion molecular clouds using VLA \citep{Tobin20}. The assumption of optically thin emission for calculating disk dust mass is more accurate at longer wavelengths and therefore VLA measurements provide better mass estimates. Restricting our sample to disks that are classified as Class I, have measured disk masses and protostellar luminosities, we end up with a sample size of 58. 

However, due to the more limited sensitivity of the VLA observations, their $3 \sigma$ detection limit corresponds to $\sim 20 M_{\oplus}$ for a $0.1 L_{\odot}$ star. The shape of the disk mass CDFs for the Class I VLA sample and the Class I ALMA sample therefore cannot be matched through a constant scaling factor as the ALMA sample contains many low mass disks and therefore is significantly more bottom-heavy. Both samples can only be compared for the highest mass disks, where the two samples are more likely to be complete. Therefore, to estimate the multiplicative factor by which we should scale up the Class II disk masses from \cite{Manara22}, we match the $3 \sigma$ (99.7th percentile) values of the disk masses from the two surveys, which gives a factor of $\sim 3$. Our result is in agreement with \cite{Tychoniec2020}, who found that Class I disks are at least 3 times more massive than Class II disks in various star forming regions.

We have implicitly assumed that the scaling factor for disk masses between the Class I and II stages is the same for the entire range of stellar masses. However, if the $M_{\rm disk} - M_\star$ relation steepens with time \citep[there is some evidence for this, see][]{Pascucci2016, Ansdell2017, Manara22}, then the multiplicative factor ought to be larger for small $M_\star$. Since Class I disks are enshrouded by protostellar envelopes that contribute to the total luminosities, $M_\star$ is typically not known for Class I disks (which is why we use Class II disks to calculate our CDFs). We can crudely relate protostellar luminosities to stellar masses and apply a $M_\star$-dependent multiplicative factor to gauge the effect on our results.

\begin{figure*}
    \includegraphics[width=\linewidth]{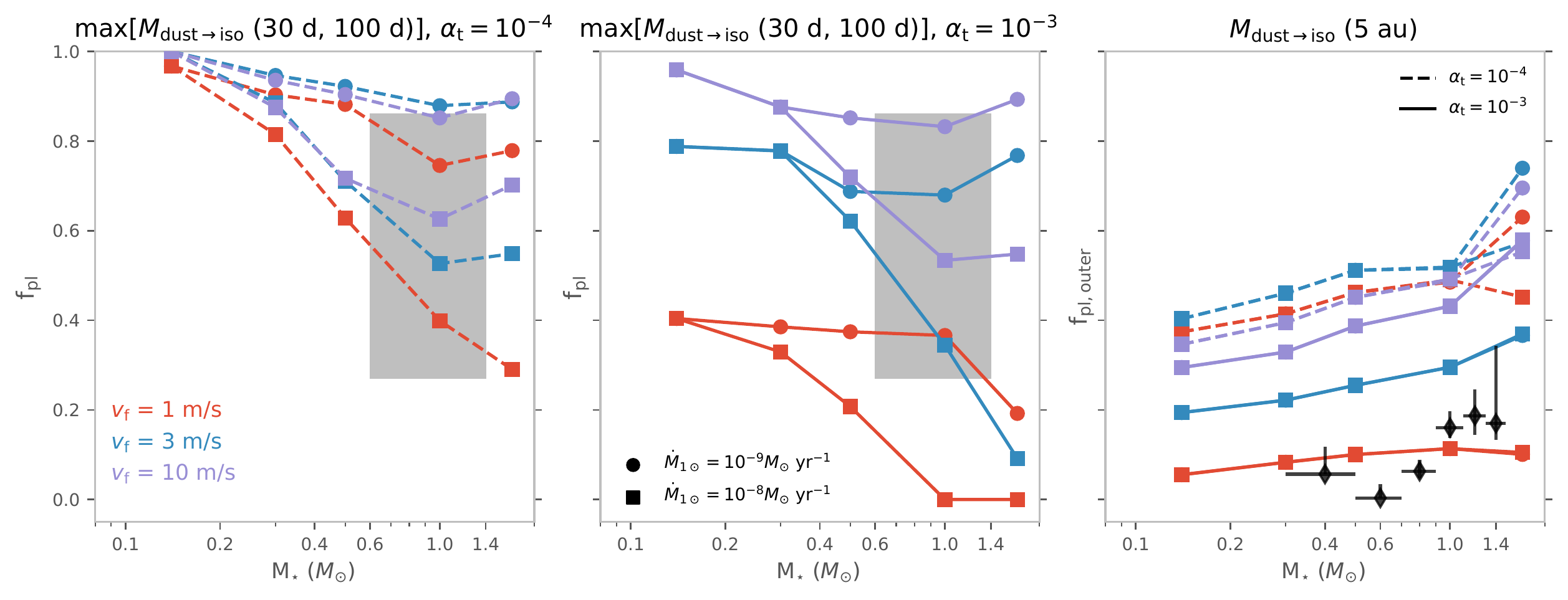}
    \includegraphics[width=\linewidth]{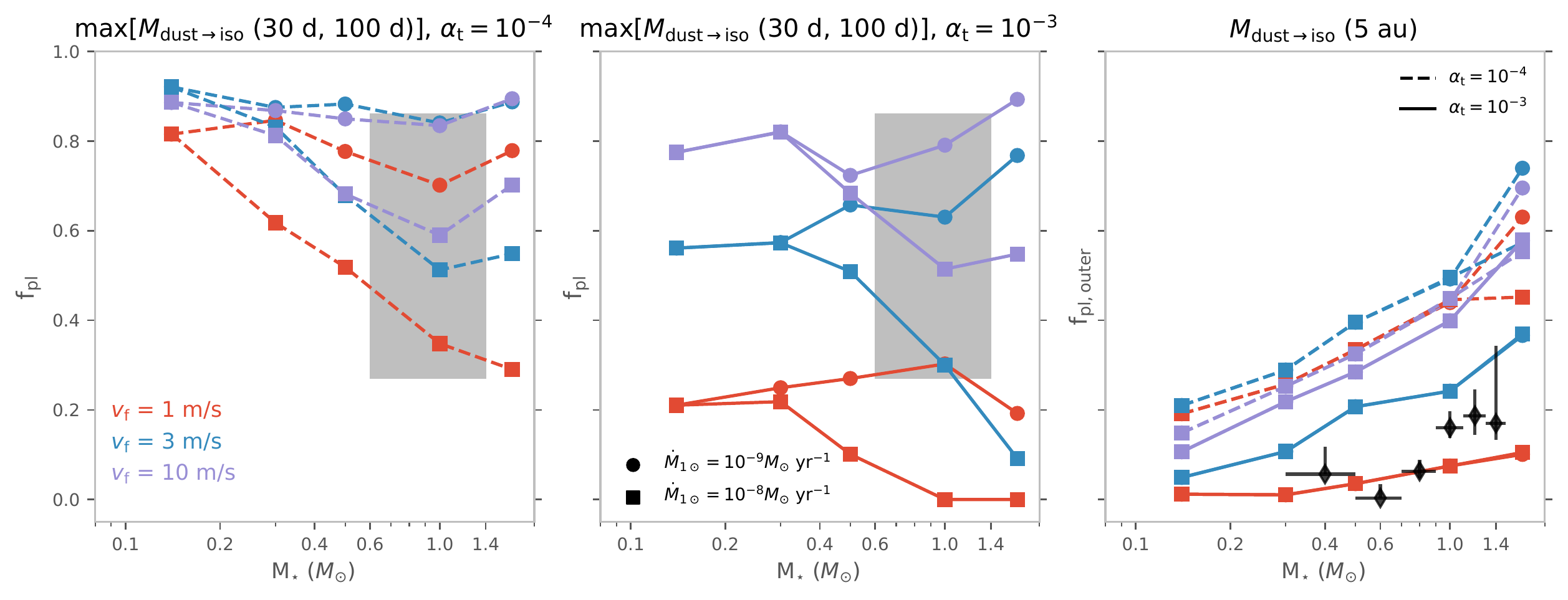}
    \caption{Fraction of disks that can create isolation masses when the disk mass distribution of \citet{Manara22} are scaled to \citet{Tobin20} in $M_\star$-dependent manner (see text for more detail). The left and the middle columns illustrate $f_{\rm pl,inner}$, equivalent to the right two panels of Figure~\ref{fig:fdisk_fpl} while the right column shows $f_{\rm pl,outer}$, equivalent to the middle panel of Figure~\ref{fig:Jup_cnxn}.The top and bottom rows correspond to a $M_\star$-bin dependent scaling of the \cite{Manara22} disk masses assuming $M_{\rm dust} \propto M_\star^{0.125}$ and $M_{\rm dust} \propto M_\star^{0.5}$ respectively. These results are in qualitative agreement with our conclusions that $f_{\rm pl}$ decreases with $M_\star$ for super-Earths and increases with $M_\star$ for giant planets. Although, the turnover in $f_{\rm pl}$ for $M_\star \lesssim 0.4 M_\odot$ is no longer present, the isolation mass of cores in the inner disk around lowest mass stars is small and mergers of these cores likely only produce super-Earths and not mini-Neptunes.}
    \label{fig:Mstar_scale_Mdust}
\end{figure*}

Linear regression on the Class I VLA data from \cite{Tobin20} yields $M_{\rm dust} \propto L_{\rm bol}^{0.25}$. If $L_{\rm bol}$ is primarily set by the pre-main sequence protostar's luminosity (i.e. $L_{\rm bol} \propto M_\star^{1.5}$, as assumed in \S~\ref{sec:theory}), then $M_{\rm dust} \propto M_\star^{0.375}$. If $L_{\rm bol}$ is set by accretion luminosity $G \dot{M}_\star M_\star / R$ instead, the two limiting power law indices for the $L_{\rm bol} - M_\star$ relation are (see \citealt{Dunham2014} for a review): i) $L_{\rm bol} \propto M_\star^{1/2}$ for isothermal core collapse ($\dot{M}_\star \propto M_\star^0$, \citealt{Shu1977}) and $R \propto M_\star^{1/2}$ as for a pre-main sequence star (the radius-mass scaling here is derived assuming a fully convective star), and ii) $L_{\rm bol} \propto M_\star^2$ for competitive accretion ($\dot{M}_\star \propto M_{\rm i}^{2/3} M_{\rm f}$, $M_{\rm i}$ is the instantaneous mass and $M_{\rm f} \sim M_\star$ is the final mass of the protostar, \citealt{Bonnell2001, McKee2010}) and for $R \propto M_\star^0$ if $R$ is the size of the protostellar core and assumed independent of $M_\star$. These two limiting cases yield $M_{\rm dust} \propto M_\star^{0.125}$ and $M_{\rm dust} \propto M_\star^{0.5}$, both of which are significantly shallower than the $M_{\rm dust} \propto M_\star^1$ relation we get for the \cite{Manara22} sample. Figure~\ref{fig:Mstar_scale_Mdust} shows the effect of scaling disk mass in different stellar bins by $3 \times (M_{\star, {\rm bin}} / 1.7 M_\odot)^{(\beta_{\rm T} - \beta_{\rm M})}$, where $\beta_{\rm T} = 0.125$ or $0.5$ and $\beta_{\rm M} \sim 1$ correspond to the power law indices characterizing $M_{\rm dust} - M_\star$ relation for the \cite{Tobin20} and \cite{Manara22} samples respectively and $M_{\star, {\rm bin}}$ is the mid-point of the stellar mass bin. The overall effect is to boost $f_{\rm pl}$ for lower mass stars such that super-Earth occurrence rate rises even more sharply with decreasing $M_\star$. The fraction of disks $f_{\rm pl, outer}$ that form a giant planet core at 5 au still rises with $M_\star$, albeit more gradually. However, the turnover in $f_{\rm pl}$ for $M_\star \lesssim 0.4 M_\odot$ is no longer present. Regardless, the sub-Earth isolation mass of cores in the inner disk around the lowest mass stars is small and may not grow into planets detectable as mini-Neptunes. Verifying this statement would require simulating orbital instabilities of systems of these isolation masses, which is a subject of future work.

\bibliography{mdwarf}{}

\begin{thebibliography}{}
\expandafter\ifx\csname natexlab\endcsname\relax\def\natexlab#1{#1}\fi
\providecommand{\url}[1]{\href{#1}{#1}}
\providecommand{\dodoi}[1]{doi:~\href{http://doi.org/#1}{\nolinkurl{#1}}}
\providecommand{\doeprint}[1]{\href{http://ascl.net/#1}{\nolinkurl{http://ascl.net/#1}}}
\providecommand{\doarXiv}[1]{\href{https://arxiv.org/abs/#1}{\nolinkurl{https://arxiv.org/abs/#1}}}

\bibitem[{{Ansdell} {et~al.}(2017){Ansdell}, {Williams}, {Manara}, {Miotello},
  {Facchini}, {van der Marel}, {Testi}, \& {van Dishoeck}}]{Ansdell2017}
{Ansdell}, M., {Williams}, J.~P., {Manara}, C.~F., {et~al.} 2017, \aj, 153,
  240, \dodoi{10.3847/1538-3881/aa69c0}

\bibitem[{{Best} {et~al.}(2023){Best}, {Sefilian}, \& {Petrovich}}]{Best23}
{Best}, S., {Sefilian}, A.~A., \& {Petrovich}, C. 2023, arXiv e-prints,
  arXiv:2304.02045, \dodoi{10.48550/arXiv.2304.02045}

\bibitem[{{Birnstiel} {et~al.}(2011){Birnstiel}, {Ormel}, \&
  {Dullemond}}]{Birnstiel11}
{Birnstiel}, T., {Ormel}, C.~W., \& {Dullemond}, C.~P. 2011, \aap, 525, A11,
  \dodoi{10.1051/0004-6361/201015228}

\bibitem[{{Bitsch} {et~al.}(2018){Bitsch}, {Morbidelli}, {Johansen}, {Lega},
  {Lambrechts}, \& {Crida}}]{Bitsch18}
{Bitsch}, B., {Morbidelli}, A., {Johansen}, A., {et~al.} 2018, \aap, 612, A30,
  \dodoi{10.1051/0004-6361/201731931}

\bibitem[{{Bonnell} {et~al.}(2001){Bonnell}, {Bate}, {Clarke}, \&
  {Pringle}}]{Bonnell2001}
{Bonnell}, I.~A., {Bate}, M.~R., {Clarke}, C.~J., \& {Pringle}, J.~E. 2001,
  \mnras, 323, 785, \dodoi{10.1046/j.1365-8711.2001.04270.x}

\bibitem[{{Bonomo} {et~al.}(2023){Bonomo}, {Dumusque}, {Massa}, {Mortier},
  {Bongiolatti}, {Malavolta}, {Sozzetti}, {Buchhave}, {Damasso}, {Haywood},
  {Morbidelli}, {Latham}, {Molinari}, {Pepe}, {Poretti}, {Udry}, {Affer},
  {Boschin}, {Charbonneau}, {Cosentino}, {Cretignier}, {Ghedina}, {Lega},
  {L{\'o}pez-Morales}, {Margini}, {Mart{\'\i}nez Fiorenzano}, {Mayor},
  {Micela}, {Pedani}, {Pinamonti}, {Rice}, {Sasselov}, {Tronsgaard}, \&
  {Vanderburg}}]{Bonomo2023}
{Bonomo}, A.~S., {Dumusque}, X., {Massa}, A., {et~al.} 2023, arXiv e-prints,
  arXiv:2304.05773, \dodoi{10.48550/arXiv.2304.05773}

\bibitem[{{Brady} \& {Bean}(2022)}]{Brady22}
{Brady}, M.~T., \& {Bean}, J.~L. 2022, \aj, 163, 255,
  \dodoi{10.3847/1538-3881/ac64a0}

\bibitem[{{Bryan} {et~al.}(2019){Bryan}, {Knutson}, {Lee}, {Fulton}, {Batygin},
  {Ngo}, \& {Meshkat}}]{Bryan19}
{Bryan}, M.~L., {Knutson}, H.~A., {Lee}, E.~J., {et~al.} 2019, \aj, 157, 52,
  \dodoi{10.3847/1538-3881/aaf57f}

\bibitem[{{Burkhardt} {et~al.}(2021){Burkhardt}, {Spitzer}, {Morbidelli},
  {Budde}, {Render}, {Kruijer}, \& {Kleine}}]{Burkhardt21}
{Burkhardt}, C., {Spitzer}, F., {Morbidelli}, A., {et~al.} 2021, Science
  Advances, 7, eabj7601, \dodoi{10.1126/sciadv.abj7601}

\bibitem[{{Chachan} {et~al.}(2021){Chachan}, {Lee}, \& {Knutson}}]{Chachan21}
{Chachan}, Y., {Lee}, E.~J., \& {Knutson}, H.~A. 2021, \apj, 919, 63,
  \dodoi{10.3847/1538-4357/ac0bb6}

\bibitem[{{Chachan} {et~al.}(2022){Chachan}, {Dalba}, {Knutson}, {Fulton},
  {Thorngren}, {Beichman}, {Ciardi}, {Howard}, \& {Van Zandt}}]{Chachan22}
{Chachan}, Y., {Dalba}, P.~A., {Knutson}, H.~A., {et~al.} 2022, \apj, 926, 62,
  \dodoi{10.3847/1538-4357/ac3ed6}

\bibitem[{{Choi} {et~al.}(2016){Choi}, {Dotter}, {Conroy}, {Cantiello},
  {Paxton}, \& {Johnson}}]{Choi16}
{Choi}, J., {Dotter}, A., {Conroy}, C., {et~al.} 2016, \apj, 823, 102,
  \dodoi{10.3847/0004-637X/823/2/102}

\bibitem[{{Clanton} \& {Gaudi}(2014)}]{Clanton14}
{Clanton}, C., \& {Gaudi}, B.~S. 2014, \apj, 791, 91,
  \dodoi{10.1088/0004-637X/791/2/91}

\bibitem[{{D'Alessio} {et~al.}(1998){D'Alessio}, {Cant{\"o}}, {Calvet}, \&
  {Lizano}}]{Dalessio98}
{D'Alessio}, P., {Cant{\"o}}, J., {Calvet}, N., \& {Lizano}, S. 1998, \apj,
  500, 411, \dodoi{10.1086/305702}

\bibitem[{{Dawson} {et~al.}(2016){Dawson}, {Lee}, \& {Chiang}}]{Dawson16}
{Dawson}, R.~I., {Lee}, E.~J., \& {Chiang}, E. 2016, \apj, 822, 54,
  \dodoi{10.3847/0004-637X/822/1/54}

\bibitem[{{Dotter}(2016)}]{Dotter16}
{Dotter}, A. 2016, \apjs, 222, 8, \dodoi{10.3847/0067-0049/222/1/8}

\bibitem[{{Dressing} \& {Charbonneau}(2015)}]{Dressing15}
{Dressing}, C.~D., \& {Charbonneau}, D. 2015, \apj, 807, 45,
  \dodoi{10.1088/0004-637X/807/1/45}

\bibitem[{{Dr{\k{a}}{\.z}kowska} \& {Alibert}(2017)}]{Drazkowska17}
{Dr{\k{a}}{\.z}kowska}, J., \& {Alibert}, Y. 2017, \aap, 608, A92,
  \dodoi{10.1051/0004-6361/201731491}

\bibitem[{{Dunham} {et~al.}(2014){Dunham}, {Stutz}, {Allen}, {Evans},
  {Fischer}, {Megeath}, {Myers}, {Offner}, {Poteet}, {Tobin}, \&
  {Vorobyov}}]{Dunham2014}
{Dunham}, M.~M., {Stutz}, A.~M., {Allen}, L.~E., {et~al.} 2014, in Protostars
  and Planets VI, ed. H.~{Beuther}, R.~S. {Klessen}, C.~P. {Dullemond}, \&
  T.~{Henning}, 195--218, \dodoi{10.2458/azu_uapress_9780816531240-ch009}

\bibitem[{{Fulton} {et~al.}(2021){Fulton}, {Rosenthal}, {Hirsch}, {Isaacson},
  {Howard}, {Dedrick}, {Sherstyuk}, {Blunt}, {Petigura}, {Knutson}, {Behmard},
  {Chontos}, {Crepp}, {Crossfield}, {Dalba}, {Fischer}, {Henry}, {Kane},
  {Kosiarek}, {Marcy}, {Rubenzahl}, {Weiss}, \& {Wright}}]{Fulton21}
{Fulton}, B.~J., {Rosenthal}, L.~J., {Hirsch}, L.~A., {et~al.} 2021, \apjs,
  255, 14, \dodoi{10.3847/1538-4365/abfcc1}

\bibitem[{{Gaidos} {et~al.}(2016){Gaidos}, {Mann}, {Kraus}, \&
  {Ireland}}]{Gaidos2016}
{Gaidos}, E., {Mann}, A.~W., {Kraus}, A.~L., \& {Ireland}, M. 2016, \mnras,
  457, 2877, \dodoi{10.1093/mnras/stw097}

\bibitem[{{Garaud} \& {Lin}(2007)}]{Garaud07}
{Garaud}, P., \& {Lin}, D.~N.~C. 2007, \apj, 654, 606, \dodoi{10.1086/509041}

\bibitem[{{Hartmann} {et~al.}(1998){Hartmann}, {Calvet}, {Gullbring}, \&
  {D'Alessio}}]{Hartmann98}
{Hartmann}, L., {Calvet}, N., {Gullbring}, E., \& {D'Alessio}, P. 1998, \apj,
  495, 385, \dodoi{10.1086/305277}

\bibitem[{{Herman} {et~al.}(2019){Herman}, {Zhu}, \& {Wu}}]{Herman19}
{Herman}, M.~K., {Zhu}, W., \& {Wu}, Y. 2019, \aj, 157, 248,
  \dodoi{10.3847/1538-3881/ab1f70}

\bibitem[{{Hsu} {et~al.}(2020){Hsu}, {Ford}, \& {Terrien}}]{Hsu20}
{Hsu}, D.~C., {Ford}, E.~B., \& {Terrien}, R. 2020, \mnras, 498, 2249,
  \dodoi{10.1093/mnras/staa2391}

\bibitem[{{Ida} {et~al.}(2016){Ida}, {Guillot}, \& {Morbidelli}}]{Ida16}
{Ida}, S., {Guillot}, T., \& {Morbidelli}, A. 2016, \aap, 591, A72,
  \dodoi{10.1051/0004-6361/201628099}

\bibitem[{{Johansen} \& {Lambrechts}(2017)}]{Johansen17}
{Johansen}, A., \& {Lambrechts}, M. 2017, Annual Review of Earth and Planetary
  Sciences, 45, 359, \dodoi{10.1146/annurev-earth-063016-020226}

\bibitem[{{Kimura} {et~al.}(2020){Kimura}, {Wada}, {Kobayashi}, {Senshu},
  {Hirai}, {Yoshida}, {Kobayashi}, {Hong}, {Arai}, {Ishibashi}, \&
  {Yamada}}]{Kimura20}
{Kimura}, H., {Wada}, K., {Kobayashi}, H., {et~al.} 2020, \mnras, 498, 1801,
  \dodoi{10.1093/mnras/staa2467}

\bibitem[{{Lambrechts} {et~al.}(2014){Lambrechts}, {Johansen}, \&
  {Morbidelli}}]{Lambrechts14}
{Lambrechts}, M., {Johansen}, A., \& {Morbidelli}, A. 2014, \aap, 572, A35,
  \dodoi{10.1051/0004-6361/201423814}

\bibitem[{{Lambrechts} {et~al.}(2019){Lambrechts}, {Morbidelli}, {Jacobson},
  {Johansen}, {Bitsch}, {Izidoro}, \& {Raymond}}]{Lambrechts19}
{Lambrechts}, M., {Morbidelli}, A., {Jacobson}, S.~A., {et~al.} 2019, \aap,
  627, A83, \dodoi{10.1051/0004-6361/201834229}

\bibitem[{{Lee}(2019)}]{Lee19}
{Lee}, E.~J. 2019, \apj, 878, 36, \dodoi{10.3847/1538-4357/ab1b40}

\bibitem[{{Lee} \& {Chiang}(2015)}]{Lee15}
{Lee}, E.~J., \& {Chiang}, E. 2015, \apj, 811, 41,
  \dodoi{10.1088/0004-637X/811/1/41}

\bibitem[{{Lee} \& {Chiang}(2016)}]{Lee16}
---. 2016, \apj, 817, 90, \dodoi{10.3847/0004-637X/817/2/90}

\bibitem[{{Lee} \& {Connors}(2021)}]{Lee21}
{Lee}, E.~J., \& {Connors}, N.~J. 2021, \apj, 908, 32,
  \dodoi{10.3847/1538-4357/abd6c7}

\bibitem[{{Lee} {et~al.}(2022){Lee}, {Karalis}, \& {Thorngren}}]{Lee22}
{Lee}, E.~J., {Karalis}, A., \& {Thorngren}, D.~P. 2022, \apj, 941, 186,
  \dodoi{10.3847/1538-4357/ac9c66}

\bibitem[{{Lin} {et~al.}(2018){Lin}, {Lee}, \& {Chiang}}]{Lin18}
{Lin}, J.~W., {Lee}, E.~J., \& {Chiang}, E. 2018, \mnras, 480, 4338,
  \dodoi{10.1093/mnras/sty2159}

\bibitem[{{Liu} {et~al.}(2020){Liu}, {Lambrechts}, {Johansen}, {Pascucci}, \&
  {Henning}}]{Liu2020}
{Liu}, B., {Lambrechts}, M., {Johansen}, A., {Pascucci}, I., \& {Henning}, T.
  2020, \aap, 638, A88, \dodoi{10.1051/0004-6361/202037720}

\bibitem[{{Lynden-Bell} \& {Pringle}(1974)}]{Lynden-Bell74}
{Lynden-Bell}, D., \& {Pringle}, J.~E. 1974, \mnras, 168, 603,
  \dodoi{10.1093/mnras/168.3.603}

\bibitem[{{Mamajek}(2009)}]{Mamajek2009}
{Mamajek}, E.~E. 2009, in American Institute of Physics Conference Series, Vol.
  1158, Exoplanets and Disks: Their Formation and Diversity, ed. T.~{Usuda},
  M.~{Tamura}, \& M.~{Ishii}, 3--10, \dodoi{10.1063/1.3215910}

\bibitem[{{Manara} {et~al.}(2022){Manara}, {Ansdell}, {Rosotti}, {Hughes},
  {Armitage}, {Lodato}, \& {Williams}}]{Manara22}
{Manara}, C.~F., {Ansdell}, M., {Rosotti}, G.~P., {et~al.} 2022, arXiv
  e-prints, arXiv:2203.09930, \dodoi{10.48550/arXiv.2203.09930}

\bibitem[{{Manara} {et~al.}(2018){Manara}, {Morbidelli}, \&
  {Guillot}}]{Manara18}
{Manara}, C.~F., {Morbidelli}, A., \& {Guillot}, T. 2018, \aap, 618, L3,
  \dodoi{10.1051/0004-6361/201834076}

\bibitem[{{McKee} \& {Offner}(2010)}]{McKee2010}
{McKee}, C.~F., \& {Offner}, S. S.~R. 2010, \apj, 716, 167,
  \dodoi{10.1088/0004-637X/716/1/167}

\bibitem[{{Ment} \& {Charbonneau}(2023)}]{Ment23}
{Ment}, K., \& {Charbonneau}, D. 2023, arXiv e-prints, arXiv:2302.04242,
  \dodoi{10.48550/arXiv.2302.04242}

\bibitem[{{Michel} {et~al.}(2021){Michel}, {van der Marel}, \&
  {Matthews}}]{Michel21}
{Michel}, A., {van der Marel}, N., \& {Matthews}, B.~C. 2021, \apj, 921, 72,
  \dodoi{10.3847/1538-4357/ac1bbb}

\bibitem[{{Moe} \& {Kratter}(2021)}]{Moe21}
{Moe}, M., \& {Kratter}, K.~M. 2021, \mnras, 507, 3593,
  \dodoi{10.1093/mnras/stab2328}

\bibitem[{{Mulders} {et~al.}(2021){Mulders}, {Dr{\k{a}}{\.z}kowska}, {van der
  Marel}, {Ciesla}, \& {Pascucci}}]{Mulders21}
{Mulders}, G.~D., {Dr{\k{a}}{\.z}kowska}, J., {van der Marel}, N., {Ciesla},
  F.~J., \& {Pascucci}, I. 2021, \apjl, 920, L1,
  \dodoi{10.3847/2041-8213/ac2947}

\bibitem[{{Mulders} {et~al.}(2015){Mulders}, {Pascucci}, \& {Apai}}]{Mulders15}
{Mulders}, G.~D., {Pascucci}, I., \& {Apai}, D. 2015, \apj, 798, 112,
  \dodoi{10.1088/0004-637X/798/2/112}

\bibitem[{{Musiolik} \& {Wurm}(2019)}]{Musiolik19}
{Musiolik}, G., \& {Wurm}, G. 2019, \apj, 873, 58,
  \dodoi{10.3847/1538-4357/ab0428}

\bibitem[{{Najita} \& {Kenyon}(2014)}]{Najita14}
{Najita}, J.~R., \& {Kenyon}, S.~J. 2014, \mnras, 445, 3315,
  \dodoi{10.1093/mnras/stu1994}

\bibitem[{{Nakagawa} {et~al.}(1986){Nakagawa}, {Sekiya}, \&
  {Hayashi}}]{Nakagawa86}
{Nakagawa}, Y., {Sekiya}, M., \& {Hayashi}, C. 1986, \icarus, 67, 375,
  \dodoi{10.1016/0019-1035(86)90121-1}

\bibitem[{{Nielsen} {et~al.}(2019){Nielsen}, {De Rosa}, {Macintosh}, {Wang},
  {Ruffio}, {Chiang}, {Marley}, {Saumon}, {Savransky}, {Ammons}, {Bailey},
  {Barman}, {Blain}, {Bulger}, {Burrows}, {Chilcote}, {Cotten}, {Czekala},
  {Doyon}, {Duch{\^e}ne}, {Esposito}, {Fabrycky}, {Fitzgerald}, {Follette},
  {Fortney}, {Gerard}, {Goodsell}, {Graham}, {Greenbaum}, {Hibon}, {Hinkley},
  {Hirsch}, {Hom}, {Hung}, {Dawson}, {Ingraham}, {Kalas}, {Konopacky},
  {Larkin}, {Lee}, {Lin}, {Maire}, {Marchis}, {Marois}, {Metchev},
  {Millar-Blanchaer}, {Morzinski}, {Oppenheimer}, {Palmer}, {Patience},
  {Perrin}, {Poyneer}, {Pueyo}, {Rafikov}, {Rajan}, {Rameau}, {Rantakyr{\"o}},
  {Ren}, {Schneider}, {Sivaramakrishnan}, {Song}, {Soummer}, {Tallis},
  {Thomas}, {Ward-Duong}, \& {Wolff}}]{Nielsen19}
{Nielsen}, E.~L., {De Rosa}, R.~J., {Macintosh}, B., {et~al.} 2019, \aj, 158,
  13, \dodoi{10.3847/1538-3881/ab16e9}

\bibitem[{{Oka} {et~al.}(2011){Oka}, {Nakamoto}, \& {Ida}}]{Oka11}
{Oka}, A., {Nakamoto}, T., \& {Ida}, S. 2011, \apj, 738, 141,
  \dodoi{10.1088/0004-637X/738/2/141}

\bibitem[{{Ormel}(2017)}]{Ormel17}
{Ormel}, C.~W. 2017, in Astrophysics and Space Science Library, Vol. 445,
  Formation, Evolution, and Dynamics of Young Solar Systems, ed. M.~{Pessah} \&
  O.~{Gressel}, 197, \dodoi{10.1007/978-3-319-60609-5\_7}

\bibitem[{{Ormel} \& {Klahr}(2010)}]{Ormel10}
{Ormel}, C.~W., \& {Klahr}, H.~H. 2010, \aap, 520, A43,
  \dodoi{10.1051/0004-6361/201014903}

\bibitem[{{Ormel} \& {Liu}(2018)}]{Ormel18}
{Ormel}, C.~W., \& {Liu}, B. 2018, \aap, 615, A178,
  \dodoi{10.1051/0004-6361/201732562}

\bibitem[{{Pascucci} {et~al.}(2016){Pascucci}, {Testi}, {Herczeg}, {Long},
  {Manara}, {Hendler}, {Mulders}, {Krijt}, {Ciesla}, {Henning}, {Mohanty},
  {Drabek-Maunder}, {Apai}, {Sz{\H{u}}cs}, {Sacco}, \&
  {Olofsson}}]{Pascucci2016}
{Pascucci}, I., {Testi}, L., {Herczeg}, G.~J., {et~al.} 2016, \apj, 831, 125,
  \dodoi{10.3847/0004-637X/831/2/125}

\bibitem[{{Perets} \& {Murray-Clay}(2011)}]{Perets11}
{Perets}, H.~B., \& {Murray-Clay}, R.~A. 2011, \apj, 733, 56,
  \dodoi{10.1088/0004-637X/733/1/56}

\bibitem[{{Pinilla} {et~al.}(2016){Pinilla}, {Klarmann}, {Birnstiel},
  {Benisty}, {Dominik}, \& {Dullemond}}]{Pinilla16}
{Pinilla}, P., {Klarmann}, L., {Birnstiel}, T., {et~al.} 2016, \aap, 585, A35,
  \dodoi{10.1051/0004-6361/201527131}

\bibitem[{{Piso} {et~al.}(2015){Piso}, {Youdin}, \& {Murray-Clay}}]{Piso15}
{Piso}, A.-M.~A., {Youdin}, A.~N., \& {Murray-Clay}, R.~A. 2015, \apj, 800, 82,
  \dodoi{10.1088/0004-637X/800/2/82}

\bibitem[{{Raghavan} {et~al.}(2010){Raghavan}, {McAlister}, {Henry}, {Latham},
  {Marcy}, {Mason}, {Gies}, {White}, \& {ten Brummelaar}}]{Raghavan10}
{Raghavan}, D., {McAlister}, H.~A., {Henry}, T.~J., {et~al.} 2010, \apjs, 190,
  1, \dodoi{10.1088/0067-0049/190/1/1}

\bibitem[{{Rosenthal} {et~al.}(2022){Rosenthal}, {Knutson}, {Chachan}, {Dai},
  {Howard}, {Fulton}, {Chontos}, {Crepp}, {Dalba}, {Henry}, {Kane}, {Petigura},
  {Weiss}, \& {Wright}}]{Rosenthal22}
{Rosenthal}, L.~J., {Knutson}, H.~A., {Chachan}, Y., {et~al.} 2022, \apjs, 262,
  1, \dodoi{10.3847/1538-4365/ac7230}

\bibitem[{{Shu}(1977)}]{Shu1977}
{Shu}, F.~H. 1977, \apj, 214, 488, \dodoi{10.1086/155274}

\bibitem[{{Tobin} {et~al.}(2020){Tobin}, {Sheehan}, {Megeath},
  {D{\'\i}az-Rodr{\'\i}guez}, {Offner}, {Murillo}, {van 't Hoff}, {van
  Dishoeck}, {Osorio}, {Anglada}, {Furlan}, {Stutz}, {Reynolds}, {Karnath},
  {Fischer}, {Persson}, {Looney}, {Li}, {Stephens}, {Chandler}, {Cox},
  {Dunham}, {Tychoniec}, {Kama}, {Kratter}, {Kounkel}, {Mazur}, {Maud},
  {Patel}, {Perez}, {Sadavoy}, {Segura-Cox}, {Sharma}, {Stephenson}, {Watson},
  \& {Wyrowski}}]{Tobin20}
{Tobin}, J.~J., {Sheehan}, P.~D., {Megeath}, S.~T., {et~al.} 2020, \apj, 890,
  130, \dodoi{10.3847/1538-4357/ab6f64}

\bibitem[{{Tychoniec} {et~al.}(2020){Tychoniec}, {Manara}, {Rosotti}, {van
  Dishoeck}, {Cridland}, {Hsieh}, {Murillo}, {Segura-Cox}, {van Terwisga}, \&
  {Tobin}}]{Tychoniec2020}
{Tychoniec}, {\L}., {Manara}, C.~F., {Rosotti}, G.~P., {et~al.} 2020, \aap,
  640, A19, \dodoi{10.1051/0004-6361/202037851}

\bibitem[{{Van Zandt} {et~al.}(2023){Van Zandt}, {Petigura}, {MacDougall},
  {Gilbert}, {Lubin}, {Barclay}, {Batalha}, {Crossfield}, {Dressing}, {Fulton},
  {Howard}, {Huber}, {Isaacson}, {Kane}, {Robertson}, {Roy}, {Weiss},
  {Behmard}, {Beard}, {Chontos}, {Dai}, {Dalba}, {Fetherolf}, {Giacalone},
  {Henze}, {Hill}, {Hirsch}, {Holcomb}, {Howell}, {Jenkins}, {Latham}, {Mayo},
  {Mireles}, {Mo{\v{c}}nik}, {Murphy}, {Pidhorodetska}, {Polanski}, {Ricker},
  {Rosenthal}, {Rubenzahl}, {Seager}, {Scarsdale}, {Turtelboom}, {Vanderspek},
  \& {Winn}}]{VanZandt2023}
{Van Zandt}, J., {Petigura}, E.~A., {MacDougall}, M., {et~al.} 2023, \aj, 165,
  60, \dodoi{10.3847/1538-3881/aca6ef}

\bibitem[{{Venturini} {et~al.}(2015){Venturini}, {Alibert}, {Benz}, \&
  {Ikoma}}]{Venturini15}
{Venturini}, J., {Alibert}, Y., {Benz}, W., \& {Ikoma}, M. 2015, \aap, 576,
  A114, \dodoi{10.1051/0004-6361/201424008}

\bibitem[{{Warren}(2011)}]{Warren11}
{Warren}, P.~H. 2011, Earth and Planetary Science Letters, 311, 93,
  \dodoi{10.1016/j.epsl.2011.08.047}

\bibitem[{{Yang} {et~al.}(2020){Yang}, {Xie}, \& {Zhou}}]{Yang20}
{Yang}, J.-Y., {Xie}, J.-W., \& {Zhou}, J.-L. 2020, \aj, 159, 164,
  \dodoi{10.3847/1538-3881/ab7373}

\bibitem[{{Zhu}(2022)}]{Zhu2022}
{Zhu}, W. 2022, \aj, 164, 5, \dodoi{10.3847/1538-3881/ac6f59}

\bibitem[{{Zhu} {et~al.}(2018){Zhu}, {Petrovich}, {Wu}, {Dong}, \&
  {Xie}}]{Zhu18_occur}
{Zhu}, W., {Petrovich}, C., {Wu}, Y., {Dong}, S., \& {Xie}, J. 2018, \apj, 860,
  101, \dodoi{10.3847/1538-4357/aac6d5}

\bibitem[{{Zhu} \& {Wu}(2018)}]{Zhu18}
{Zhu}, W., \& {Wu}, Y. 2018, \aj, 156, 92, \dodoi{10.3847/1538-3881/aad22a}

\end{thebibliography}
\bibliographystyle{aasjournal}

\end{document}